\documentclass[11pt,a4paper]{article}
\usepackage[utf8]{inputenc}
\usepackage[T1]{fontenc}
\usepackage{mathptmx}
\usepackage[english]{babel}
\usepackage[margin=2.3cm]{geometry}
\usepackage{microtype}
\usepackage{amsmath,amssymb}
\usepackage{graphicx}
\usepackage{booktabs}
\usepackage{tabularx}
\usepackage{array}
\usepackage{textcomp}
\usepackage[labelfont=bf,labelsep=period,font=small]{caption}
\usepackage[numbers,sort&compress]{natbib}
\usepackage{authblk}
\usepackage{xurl}
\usepackage[hidelinks]{hyperref}
\usepackage{setspace}
\hypersetup{
  pdftitle={Foundation-model-based multi-label phenotyping of combined hyperkinetic movement disorders},
  pdfauthor={Laura Cif, Xavier Vasques, et al.},
  pdfkeywords={hyperkinetic movement disorders; diagnosis; computational phenotyping; foundation models; Segment Anything Model 3; TabICLv2 tabular model}
}

\title{\textbf{Foundation-model-based multi-label phenotyping of combined hyperkinetic movement disorders}}
\author[1,2,\#]{Laura~Cif}
\author[2,3,*]{Zohra~Souei}
\author[2,4,*]{Diane~Demailly}
\author[1]{Mayté~Castro-Jimenez}
\author[5,6,7]{Juan~Dario~Ortigoza-Escobar}
\author[8]{Muhammad~Mushhood~Ur~Rehman}
\author[9]{Morgan~Dornadic}
\author[9]{Sophie~Huby}
\author[10]{Gun-Marie~Hariz}
\author[1]{Cécile~Hubsch}
\author[11]{Nathalie~Dorison}
\author[12,13]{Eduardo~M.~Moraud}
\author[12,14]{Jocelyne~Bloch}
\author[15]{Gabriella~Horvàth}
\author[16,17,18]{Olivier~Oullier}
\author[2]{Xavier~Vasques}
\affil[1]{Service of Neurology, Department of Clinical Neurosciences, Lausanne University Hospital (CHUV) and University of Lausanne (UNIL), Lausanne, Switzerland}
\affil[2]{Institut du Neurone, Montferrier sur Lez, France}
\affil[3]{Department of Neurosurgery, Military University Hospital of Sfax, Sfax, Tunisia}
\affil[4]{Department of Neurology, Clinique Beau Soleil, Institut Mutualiste Montpelliérain, Montpellier, France}
\affil[5]{Movement Disorders Unit, Pediatric Neurology Department, Institut de Recerca, Hospital Sant Joan de Déu, Barcelona, Spain}
\affil[6]{European Reference Network for Rare Neurological Diseases (ERN-RND), Barcelona, Spain}
\affil[7]{U-703 Centre for Biomedical Research on Rare Diseases (CIBER-ER), Instituto de Salud Carlos III, Barcelona, Spain}
\affil[8]{Edinburgh Medical School, University of Edinburgh, Scotland, UK}
\affil[9]{Department of Neurology, CHU Montpellier, 34295, Montpellier, France}
\affil[10]{Department of Clinical Neuroscience, Umeå University, Umeå, Sweden}
\affil[11]{Pediatric Neurosurgery Department, CCMR Neurogenetique, European Reference Network Brainteam Member, Rothschild Foundation Hospital, Paris, France}
\affil[12]{Defitech Center for Interventional Neurotherapies (NeuroRestore), University Hospital Lausanne and Ecole Polytechnique Fédérale de Lausanne, Lausanne, Switzerland}
\affil[13]{Neuro-X Institute, Ecole Polytechnique Fédérale de Lausanne, Geneva, Switzerland}
\affil[14]{Department of Neurosurgery, Lausanne University Hospital (CHUV) and University of Lausanne (UNIL), Lausanne, Switzerland}
\affil[15]{Department of Pediatrics, British Columbia Children's Hospital, Vancouver, British Columbia, Canada}
\affil[16]{Computing and Mathematical Sciences Division, Mohamed bin Zayed University of Artificial Intelligence (MBZUAI), Abu Dhabi, UAE}
\affil[17]{Inclusive Brains, Marseille, France}
\affil[18]{Institute for Artificial Intelligence, Biotech Dental Group, Salon-de-Provence, France}
\affil[ ]{\textsuperscript{\#}Corresponding author: Laura Cif, \href{mailto:lauracif@institutduneurone.fr}{lauracif@institutduneurone.fr}}
\affil[ ]{\textsuperscript{*}These authors contributed equally to this work}
\date{}

\begin{document}
\maketitle

\begin{abstract}
\noindent Movement disorders (MDs) frequently co-occur, yet phenomenological and severity assessment shows substantial inter-rater variability. Markerless video could improve reproducibility, but prior work is largely single-symptom, depends on standardized acquisition, and lacks validation and transfer across ages and sites.

We combined two foundation models into one frozen backbone: Segment Anything Model 3 (SAM~3) for dense, per-frame markerless segmentation summarized into geometric, contour and grid kinematic signals, and TabICLv2, a tabular foundation model, for in-context multi-label classification of eight hyperkinetic MD phenomenologies. Trained on standardized recordings of 21 adults and 4 controls, it transferred unchanged to two independent datasets, pediatric (n=12) and tremor-dominant adult (n=20), assessed with Combined Dystonia Scale for Assessment of the Motor Phenotype (CODY-SAMP); only the patient-level decision step was recalibrated per site.

Transfer succeeded through decision-step recalibration alone. Under clinician consensus labels, false positives fell to zero in both datasets. Dystonia recovered perfectly (7/7 pediatric; 15/15 adult held-out), chorea fully in children (3/3), and tremor was recovered in adults (11/15) once a tremor-rich cohort made it evaluable, through recalibration alone. Per-region effect-size analysis gave clinically coherent, phenomenology-specific signals and identified myoclonus as the principal failure. Against YOLOv8 sparse keypoints, the dense representation matched under clinician permissive labels (Jaccard 0.63 vs 0.63) and was markedly more robust under clinician-label consensus (0.93 vs 0.76).

Despite limited dataset size and composition, this frozen foundation-model backbone with light per-site calibration yields transferable, interpretable, conservative multi-label phenotyping of co-occurring hyperkinetic MDs across ages and from standardized to routine video, adding robustness on high-confidence, clinician-agreed labels. Prospective multi-centre validation is required before clinical use.

\end{abstract}

\section{Introduction}

Diagnostic methods for movement disorders (MDs) typically rely on the subjective assessment of motor symptoms, which poses inherent challenges, compounded by their etiological and phenomenological heterogeneity. Hyperkinetic MDs (HMDs, i.e., dystonia, chorea, tremor, myoclonus, athetosis, ballismus, stereotypies and tics) are defined by excessive involuntary movements~\cite{ref1} as opposed to hypokinetic MDs, with Parkinson's disease (PD), the most frequent and studied with  clinical diagnostic criteria, including bradykinesia, together with at least one of rigidity, rest tremor or postural instability, according to the United Kingdom Parkinson's Disease Society Brain Bank Criteria~\cite{ref2}.

Beyond this definitional contrast, HMDs and PD occupy very different positions in clinical research. PD is frequent, diagnostically anchored by consensus criteria and quantified by the gold standard, widely adopted instrument, the Movement Disorder Society-Unified Parkinson's Disease Rating Scale (MDS-UPDRS), which has made it the natural testbed for objective motor assessment~\cite{ref3,ref4}. HMDs are individually rare, etiologically heterogeneous, span the entire lifespan, and are assessed with a fragmented set of phenomenology-specific scales, none of which was designed for the combined presentations that are the rule rather than the exception in genetic, neurometabolic and developmental disorders. The practical consequence is a long diagnostic pathway: in cervical dystonia, arguably the most recognizable HMD and the most frequent isolated focal dystonia, the mean interval from symptom onset to diagnosis approaches seven years~\cite{ref5}.

In combined HMDs the phenotype is itself the entry point to the diagnosis. Overlapping symptoms across diseases can complicate both early diagnosis and disease monitoring~\cite{ref6}. A patient with a combined MD may, in a single examination, show dystonia, choreic flow and myoclonic jerks at once, and which of these is present, and how prominent, steers the genetic work-up, drug selection and candidacy for deep brain stimulation (DBS)~\cite{ref1,ref7,ref8}. Accurate and timely diagnosis as well as the continuous monitoring of these disorders are crucial for facilitating effective patient care.

Several features make phenotyping, the first step toward diagnosis, hard. First, these phenomenologies coexist, prominence fluctuates over time and across body regions, so phenotyping is intrinsically a multi-label problem, not a single-diagnosis classification. Second, their visual assessment carries substantial inter-rater variability even among experienced neurologists, which limits reproducibility and complicates multi-center research~\cite{ref9,ref10,ref11}. Third, the boundaries between phenomenologies require further refinement, as the pathophysiology underlying them awaits fuller understanding. This variability is not uniform across signs: agreement is high for well-defined, isolated phenomenologies such as essential tremor and dystonia, falls for myoclonus and for choreoathetotic phenomenology, and falls further in complex cases in which several signs co-occur in the same patient~\cite{ref12,ref13}. Any automated system must therefore be evaluated against labels of stated reliability rather than against an assumed gold standard.

\subsection{State of the art and research contributions}

Automatic two-dimensional (2D) and three-dimensional (3D) video analysis with deep learning (DL) was proposed over the last decade in MDs analysis focusing mostly on specific PD symptom quantification, gait disorder analysis, tremor assessment, and only marginally addressing HMDs.

\subsubsection{Computer vision in Parkinson's disease: a more mature, reference field}

In PD, markerless computer vision and human pose estimation have been deployed across the whole clinical pipeline: screening and discrimination from healthy controls, identification of individual motor signs, quantification of severity against the MDS-UPDRS, and monitoring of the response to treatment, including levodopa-induced dyskinesia, motor fluctuations and DBS~\cite{ref14,ref15,ref16,ref17,ref18}. Bradykinesia has become the exemplar of this effort, because it is the cardinal criterion for the diagnosis of PD and because the MDS-UPDRS items that capture it, finger tapping, hand movements and pronation-supination, are brief, highly standardized and easy to film. A substantial body of work now estimates these items automatically from consumer video and reports amplitude, frequency, decrement and hesitation measures that track clinician ratings and, longitudinally, disease progression, so that vision-derived bradykinesia has come to serve as a proxy for disease severity and its evolution~\cite{ref2,ref19,ref20}. Hand pose estimation has been consolidated into a dedicated methodological literature. Amprimo et al.~\cite{ref21} conducted a narrative review on the deep-learning-based hand-tracking approaches for video-based PD assessment and showed that a small number of frameworks, such as OpenPose~\cite{ref22}, DeepLabCut~\cite{ref23}, and MediaPipe~\cite{ref24}, along with custom architectures, support objective, video-based and increasingly remote quantification of bradykinesia, while identifying standardization and clinical validation as key remaining obstacles. PD therefore benefits from three converging assets that HMDs lack: a single dominant diagnosis, a universally accepted severity anchor, and short, standardized, camera-friendly tasks. However, as recently highlighted in the systematic review by Tang et al.~\cite{ref6}, the use of DL and computer vision is still limited for hyperkinetic MDs, including tic quantification and assessment of focal dystonia~\cite{ref25,ref26}. Possible explanations are the heterogeneity of their etiology, the complexity of symptom associations, body distribution, definition of tasks~\cite{ref12} that allow their optimal assessment together with a key challenge, the reliability of the ground truth for certain tasks, particularly regarding symptom recognition and rating scores.

\subsubsection{Hyperkinetic movement disorders: a far larger and almost unexplored target}

The marked PD-centricity of the field was recently confirmed by Pecoraro et al.~\cite{ref27}: of 71 studies included in their systematic review of computer-vision technologies in movement disorders, 74.6\% focused on PD, whereas only nine addressed dystonia syndromes, four essential tremor, and one Tourette syndrome; no studies investigated chorea or ballism, and none examined combined movement disorders~\cite{ref27}. Reviews of video-based, as well as artificial intelligence approaches more broadly, reach a similar conclusion: the field remains predominantly focused on PD, cohorts are generally small and single-centre, and non-parkinsonian movement phenomenologies remain largely unaddressed~\cite{ref6,ref28,ref29}.

What does exist is confined to isolated phenomenologies examined in dedicated, task-specific settings. Motor facial tics have been detected and counted automatically from video in Tourette syndrome~\cite{ref25,ref26,ref30,ref31}; focal, cervical dystonia has been quantified through head-pose and head-tremor analysis, including multi-centre deep-learning studies of head movement dynamics~\cite{ref32}; myoclonus severity has been estimated from recordings acquired according to the standardized Unified Myoclonus Rating Scale protocol~\cite{ref33}; and essential tremor has been characterized with validated video-based algorithms~\cite{ref16}. Chorea, by contrast, has been approached almost exclusively with wearable sensors rather than video, and athetosis, ballismus and stereotypies have no dedicated computer-vision literature at all.

Several reasons account for this gap, and they are specific to hyperkinetic phenomenology. HMDs are etiologically heterogeneous, so the same sign may arise from monogenic, acquired or developmental causes with different age-dependent expression. They rarely occur alone: in genetic and developmental syndromes, dystonia, chorea, myoclonus and tremor coexist within the same patient and often within the same body segment, which breaks the single-label formulation on which nearly all existing pipelines rest. Their body distribution is variable and frequently asymmetric, so the informative region differs from patient to patient. Clinical examination tasks are not uniformly standardized, and task selection can substantially influence whether, and how clearly, a movement disorder sign is elicited. Sciacca et al. demonstrated substantial task-dependent variation in the visibility of essential tremor, dystonia, cortical myoclonus, and myoclonus-dystonia across 21 clinical tasks, with inter-rater agreement varying between phenotypes and only three tasks providing the highest visibility across all four groups~\cite{ref12}. Finally, the clinical reference standard itself is subject to observer uncertainty, with reliability varying across phenomenologies and becoming more challenging in complex presentations~\cite{ref12,ref13}. Phenomenological labels are therefore best treated as graded clinical opinion rather than as an incontestable reference, which is the reasoning behind the agreement ladder we adopted for this paper.

\subsubsection{Artificial intelligence for the classification of hyperkinetic movement disorders}

Attempts to move beyond quantification of individual motor signs toward classification across hyperkinetic phenomenologies remain comparatively limited. The Next Move in Movement Disorders (NEMO) programme represents a closely related prior effort: it was designed explicitly to develop a computer-aided classification tool for HMDs by integrating clinical information, electromyography, accelerometry, and three-dimensional video acquired during a standardized experimental protocol. Expert-based phenotype classification served as the reference standard, with each case independently assessed by three experienced clinical experts and agreement of at least two required for inclusion. The protocol encompassed prespecified single-phenotype groups, initially dystonia, myoclonus, and tremor, together with a limited set of prespecified mixed-phenotype groups, including myoclonus-dystonia and dystonic tremor~\cite{ref34}. NEMO therefore established a framework for automated cross-phenotype classification, but within a multimodal laboratory setting and a predefined set of phenotype categories rather than unrestricted multi-label combinations of signs. Existing reviews reinforce this picture, with most studies focusing on individual conditions and relying on relatively small, often single-centre cohorts~\cite{ref28}.

Markerless video analysis is an attractive route to more objective, scalable assessment, requiring only a consumer camera. But prior work shares several recurring limitations: it relies on sparse pose (a few skeletal keypoints) that discards the continuous postural deformation of dystonia and chorea and may not capture small amplitude brisk, focally distributed phenomenology; it targets a single phenomenology, most often assessment of bradykinesia~\cite{ref2} as a proxy for PD severity and progression or a single body segment; and it is validated in one population under controlled acquisition, without demonstrated transfer across ages, etiologies and recording conditions~\cite{ref14,ref15,ref23,ref29,ref30,ref34,ref35,ref36,ref37}. Decisively for the present purpose, prior work is almost invariably formulated as a single-label problem, and therefore cannot report the association of signs on which the syndromic diagnosis of a combined HMD actually rests.

Foundation models can address the first two obstacles. Segment Anything Model 3 (SAM~3) provides promptable, open-vocabulary segmentation that can be used to isolate a human body densely on every frame, yielding a far richer, continuously deforming body description than a skeleton~\cite{ref38,ref39}. Tabular foundation models such as TabICLv2 classify tabular data in context, predicting from a set of labelled examples without per-dataset gradient training, which suits the small-sample, many-feature regime of clinical movement-disorder datasets~\cite{ref40,ref41}. Pairing a dense visual representation with a tabular foundation-model classifier is, to our knowledge, a new strategy for multi-label phenotyping of combined MDs.

In previous studies, we first established feasibility of multi-label phenotyping from sparse pose in a single center, followed by the demonstration of adult-to-pediatric transfer~\cite{ref36,ref37}. In parallel, we systematically reviewed the use of artificial intelligence in deep brain stimulation for movement disorders and assessed its technology readiness, identifying objective symptom quantification as the application closest to clinical translation while showing that most existing pipelines remain narrowly trained, single-centre, and reliant on task- or cohort-specific retraining~\cite{ref35}. Together, these findings motivate the present study. Sparse keypoint representations and dataset-specific training impose two important limitations: keypoints provide only a coarse representation of body configuration and may omit the continuous postural and shape information relevant to phenomenologies such as dystonia and chorea, while models that require retraining for each cohort are inherently difficult to transport across sites. Foundation models offer a route to overcoming both constraints. A frozen, general-purpose vision backbone can provide dense representations without cohort-specific end-to-end training, while a tabular in-context learner can adapt to a new site from only a small set of labelled examples, an operating regime well suited to the limited sample sizes typical of rare-disease datasets~\cite{ref39,ref40,ref41,ref42}.

Here we introduce SAM~3 dense segmentation and evaluate it on a wider set of datasets, including an adult tremor-predominant population. The central design principle is a \emph{train-once, calibrate-then-deploy} strategy: a shared predictive backbone is trained a single time on standardized examinations, and when it is applied to a new site only the final patient-level decision step, the aggregation of per-window probabilities and the decision thresholds, is recalibrated, using a small number of locally selected patients. The backbone itself is never retrained. This mirrors the practical reality of clinical deployment, where a new center can annotate a handful of patients but cannot re-run a full training procedure, and it isolates the question of transferability from that of model capacity.

We make four contributions: first use of a dense markerless representation coupled to a tabular foundation model for multi-label classification of combined MDs; demonstration of single-backbone transfer across markedly different age ranges and acquisition conditions with calibration-only adaptation; an exhaustive characterization of how calibration-patient choice affects performance; and a per-phenomenology interpretability analysis that accounts for both successes and failures. We also quantify how the dense representation compares with its sparse-keypoint predecessor as the label definition is tightened toward clinical consensus. To our knowledge, this is the first study to use a foundation-model-based computer-vision pipeline for multi-label recognition of co-occurring hyperkinetic movement phenomenologies from routine clinical video and to demonstrate transfer across heterogeneous, demographically distinct cohorts without dataset-specific retraining.

\section{Methods}

\subsection{Study design, participants, video acquisition and ethics}

\begin{figure}[tbp]
\centering
\includegraphics[width=\textwidth,trim=16 5 18 22,clip]{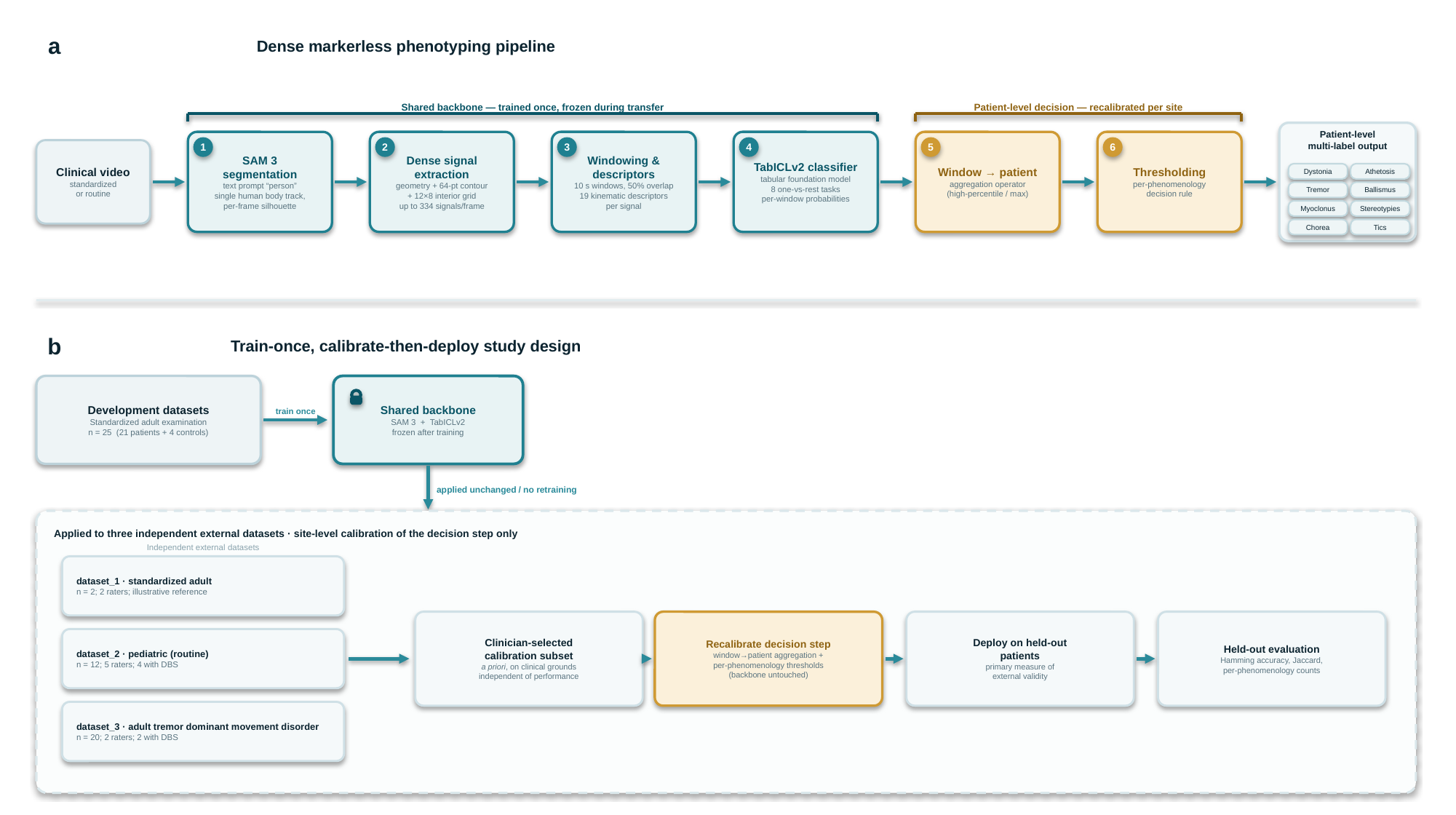}
\caption{\textbf{Transferable foundation-model phenotyping of hyperkinetic movement disorders from clinical videos.} We present a video-based framework that produces patient-level, multi-label phenotyping of eight HMDs phenomenologies (dystonia, tremor, myoclonus, chorea, athetosis, ballismus, stereotypies and tics) directly from clinical recordings. The approach pairs two foundation models: open-vocabulary segmentation with SAM~3, which isolates the human body on every frame and yields a dense set of kinematic signals, and a tabular foundation model, TabICLv2, which classifies each phenomenology without task-specific retraining. A shared backbone is trained once on standardized adult examinations and applied unchanged to independent adult and pediatric datasets; only the patient-level decision step is recalibrated per site.}
\label{fig:1}
\end{figure}

This study evaluated a video-based framework for patient-level multi-label phenotyping of HMDs from clinical video recordings (Figure~\ref{fig:1}). The overall workflow comprised three stages: model training on annotated standardized videos, external inference on independent datasets, and dataset-specific calibration of the final patient-level decision step.

\subsubsection{Training cohort}

The training dataset was collected at Beau Soleil Clinic, Montpellier, France, and included 25 participants in total: 21 patients with HMDs and 4 healthy controls. The 21 patients (12 female; mean age 46.9 $\pm$ 21.0 years, median 51 years, range 17--75 years) all carried a confirmed clinical diagnosis of HMD, with a heterogeneous etiological distribution including monogenic, acquired and developmental causes, as well as combined MD syndromes without an established cause. The four healthy controls (2 female; mean age 50.7 $\pm$ 27.7 years) had no personal or family history of movement disorders and underwent the same standardized examination protocol.

Phenomenology was characterized directly from the standardized video examinations, in which each of the eight target HMDs was coded window by window by two raters; a phenomenology was counted present for a patient when both raters scored at least one 10-second window positive. On this basis, the dataset was dominated by dystonia (21/21 patients; 100\%), followed by tremor and myoclonus (13/21 each; 62\%), athetosis (7/21; 33\%), chorea (4/21; 19\%), ballismus and tics (2/21 each; 10\%) and stereotypies (1/21; 5\%). This distribution defined the supervised signal available for model training and explains the more limited learning support for the least represented phenomenologies.

\subsubsection{Clinical assessment and video acquisition}

All participants underwent a standardized video-recorded examination (Supplementary Material S1) and were clinically assessed using Combined Dystonia Scale for Assessment of the Motor Phenotype (CODY-SAMP), a structured clinical scale developed for the assessment of combined MDs across pediatric and adult populations~\cite{ref43}. Each full examination video was rated independently by two clinicians, yielding 50 full-examination ratings. In addition, the same 25 participants contributed condition-specific videos corresponding to rest, posture and action tasks, yielding 75 additional video-derived files. Altogether, 125 video-derived files were used for annotation, feature extraction and downstream model development. Rest and posture recordings lasted 10 seconds each, whereas action recordings lasted 20 seconds. Video sessions lasted approximately 12.5 minutes when all tasks were completed, and cooperation was adequate.

\subsubsection{Evaluation datasets}

Evaluation was then performed on three independent datasets representing different clinical conditions of use (Table~\ref{tab:1}).

\begin{table}[tbp]
\centering
\caption{\textbf{Training and evaluation cohorts.} Prevalences for dataset\_2 and dataset\_3 are at the at-least-one-rater (or per-window annotation) layer. The three datasets occupy non-overlapping age ranges. Deep brain stimulation systems were active in 4/12 (dataset\_2) and 2/20 (dataset\_3) patients.}
\label{tab:1}
\scriptsize
\renewcommand{\arraystretch}{1.25}
\begin{tabularx}{\textwidth}{@{}>{\raggedright\arraybackslash}p{1.35cm}>{\raggedright\arraybackslash}p{1.25cm}>{\raggedright\arraybackslash}p{1.55cm}c>{\raggedright\arraybackslash}p{1.45cm}>{\raggedright\arraybackslash}p{2.4cm}>{\raggedright\arraybackslash}X@{}}
\toprule
\textbf{Dataset} & \textbf{n} & \textbf{Age, years (mean $\pm$ SD; range)} & \textbf{Raters} & \textbf{Acquisition} & \textbf{Dominant phenomenologies} & \textbf{Etiologies} \\
\midrule
\textbf{Training} & 25 (21 + 4 ctrl) & 46.9 $\pm$ 21.0 (17--75) & 2 & Standardized & Dystonia (100\%), tremor and myoclonus (62\%) &
\textbf{Genetic, confirmed (6/21, 29\%):} \emph{DIP2B} (n=2), \emph{LRRK2} (n=1), \emph{ADCY5} (n=1), \emph{TTPA}/AVED (n=1), \emph{TIMM8} (n=1).
\textbf{Idiopathic/unresolved (15/21, 71\%):} combined dystonia (n=4), dystonia (n=3), chorea-dystonia (n=2), chorea and ataxia (n=1), tremor-prominent disorder (n=1), myoclonus-dystonia (n=1), hypoxic cerebral palsy (n=1), neurodevelopmental disorder + tardive dystonia (n=1), hyperkinetic movement disorder, unspecified (n=1) \\
\addlinespace
\textbf{dataset\_1} & 2 & 48 and 80 & 2 & Standardized & Illustrative only (dystonia, tremor) &
Genetic generalized dystonia (n=1); unresolved combined movement disorder (n=1) \\
\addlinespace
\textbf{dataset\_2 (pediatric)} & 12 & 11.9 $\pm$ 2.6 (7.0--15.5) & 5 & Routine & Dystonia (100\%), athetosis (42\%), chorea and myoclonus (33\%) &
\emph{ADCY5}-related disorder (n=4); \emph{PANK2}-related disorder (n=3); \emph{SGCE} (sarcoglycan epsilon)-related disorder (n=2); \emph{KMT2B}-related disorder (n=2); \emph{GNAO1}-related disorder (n=1) \\
\addlinespace
\textbf{dataset\_3 (adult)} & 20 & 67.2 $\pm$ 18.5 (22--89) & 2 & Standardized (tremor-focused) & Dystonia (95\%), tremor (100\%); parkinsonism, myoclonus and chorea associated &
Parkinson's disease (n=7); essential tremor/essential tremor-plus (n=4); cervical dystonia (n=2); iatrogenic tremor (n=2); dystonic tremor (n=1); generalized dystonia (n=1); post-neuroleptic dystonia-parkinsonism (n=1); combined chorea-dystonia (n=1); anti-MAG tremor (n=1) \\
\bottomrule
\end{tabularx}
\end{table}

\textbf{Dataset\_1} included 2 additional adult patients (1 female; aged 48 and 80 years) examined under the same standardized protocol as the training dataset: one with genetic generalized dystonia and one with unresolved combined MD. It is reported here as an illustrative standardized-acquisition reference rather than as a powered evaluation cohort.

\textbf{Dataset\_2} (external) consisted of 12 pediatric patients (9 female; mean age 11.9 $\pm$ 2.6 years, range 7.0--15.5 years), all with a confirmed monogenic MD syndrome; 4 of 12 patients carried active DBS systems at the time of recording. Dataset\_2 was recorded during routine clinical practice outside the standardized protocol and therefore represented a more difficult real-world setting, with greater variability in framing, camera positioning and overall video quality. This distinction was important because the study aimed not only to evaluate performance in standardized recordings, but also to test whether the framework could remain informative when applied to suboptimal real-world clinical videos.

\textbf{Dataset\_3} consisted of 20 patients (7 male; mean age 67.2 $\pm$ 18.5 years, median 70.5 years, range 22--89 years) presenting predominantly with adult-onset movement disorders including Parkinson's disease, tremor and dystonia. Beyond the tremor and dystonia targeted by the recording protocol, parkinsonism, myoclonus and chorea were also present. Two out of 20 patients carried active DBS. The patients were examined following a short, standardized video recording protocol focusing on the assessment and recording of upper limb and cephalic tremor with kinematic motion trackers.

The evaluation datasets covered age ranges distinct from that of the training dataset (training: 17--75 years; dataset\_2: 7.0--15.5 years; dataset\_3: 22--89 years). The pediatric dataset\_2 lay entirely outside the training range, whereas dataset\_3 extended it into older age (median 70.5 years), so that both evaluation datasets represented demographic regimes shifted away from the training population. For the evaluation datasets, the framework generated patient-level multi-label predictions and, when specified, applied local calibration of the final decision step.

Although the CODY-SAMP clinical instrument includes a broader clinical assessment of movement disorder features, the present model focused specifically on the eight HMD phenomenologies. Other clinically relevant items captured such as parkinsonism, ataxia or dysarthria, were not modeled in the present study.

\subsubsection{Ethics}

Ethics approval was obtained from Beau Soleil Clinic, Montpellier, France (CESSRESS 22075132 Bis and CNIL n\textdegree{}2238428) and the University Hospital Montpellier, France (IRB-MTP\_2020\_09\_202000565). Informed consent for study participation and video recording was obtained from all patients or their legal guardians. All procedures complied with the Declaration of Helsinki, 1975, as revised in 2024.

\subsection{Annotation and label definitions}

Each video was annotated by clinicians for the presence of the eight target phenomenologies. Because the number of raters differed across datasets (Table~\ref{tab:1}), we adopted a rating-adaptive scheme for deriving patient-level reference labels, so that the same conceptual definitions could be applied consistently regardless of how many clinicians had scored a given cohort.

For a dataset scored by \emph{R} raters, a phenomenology was considered present at the patient level when at least \emph{k} of the \emph{R} raters had annotated it. Here, agreement means the number of independent raters who scored the same phenomenology as present in the same patient. We considered two options, permissive and consensus-based. For the \emph{main} one, every patient received a binary present/absent label at a given agreement threshold (a phenomenology was absent if fewer than k raters annotated it). For the \emph{restrictive} option, a phenomenology was labelled present if at least k raters annotated it and absent only if no rater annotated it; patients with intermediate, ambiguous agreement were excluded from the evaluation of that phenomenology. The consensus-based option therefore measures performance on cases with clearer clinical consensus. The agreement thresholds depended on the number of raters. For the five-rater pediatric dataset\_2, we report k $\geq$ 3/5, k $\geq$ 4/5 and k = 5/5. For the two-rater adult dataset\_3, we therefore report k $\geq$ 1/2 (at least one rater) and k = 2/2 (full consensus). The rationale for reporting this ladder rather than a single threshold is the following: clinical raters disagree on these signs, there is no single incontestable reference standard, and a model can only be judged against labels of stated reliability. The permissive option asks whether the model recovers every sign that any experienced rater considered present, including uncertain calls; the consensus one asks whether it recovers the signs that all raters independently agreed upon, that is, the part of the phenotype that is clinically unambiguous. Reading performance along this ladder therefore separates two questions that a single number conflates: whether the model tracks the reliable clinical phenotype, and whether it also reproduces borderline judgements on which clinicians themselves differ. Throughout, the consensus-based options are treated as the primary reference, because they are the ones a clinician would act upon. This choice, like the choice of calibration patients, was fixed in advance, before any held-out result was examined.

Under the at-least-one-rater option, dataset\_3 was, as expected, dominated by tremor (20/20 patients) and dystonia (19/20), with myoclonus present in a minority (5/20) and the remaining HMD phenomenologies essentially absent.

\subsection{Dense markerless representation: SAM~3 segmentation, 3D body recovery and kinematic features}

Each clinical video was processed with the Segment Anything Model 3 (SAM~3; Meta, released November 2025), used through the Ultralytics interface (version $\geq$ 8.3.237) with the semantic predictor and the text concept prompt ``person''~\cite{ref38}. SAM~3 returns one segmentation mask per detected human body on each frame. Because a clinician is frequently present in the field of view, a single human body track was maintained automatically: on the first frame the largest human body mask was selected as the patient, and on subsequent frames the patient was re-identified as the candidate mask with the highest intersection-over-union with the previous patient mask, with a centroid-distance fallback. All other persons were ignored. This yields a dense, per-frame silhouette of the patient throughout the recording.

\begin{figure}[tbp]
\centering
\includegraphics[width=0.92\textwidth]{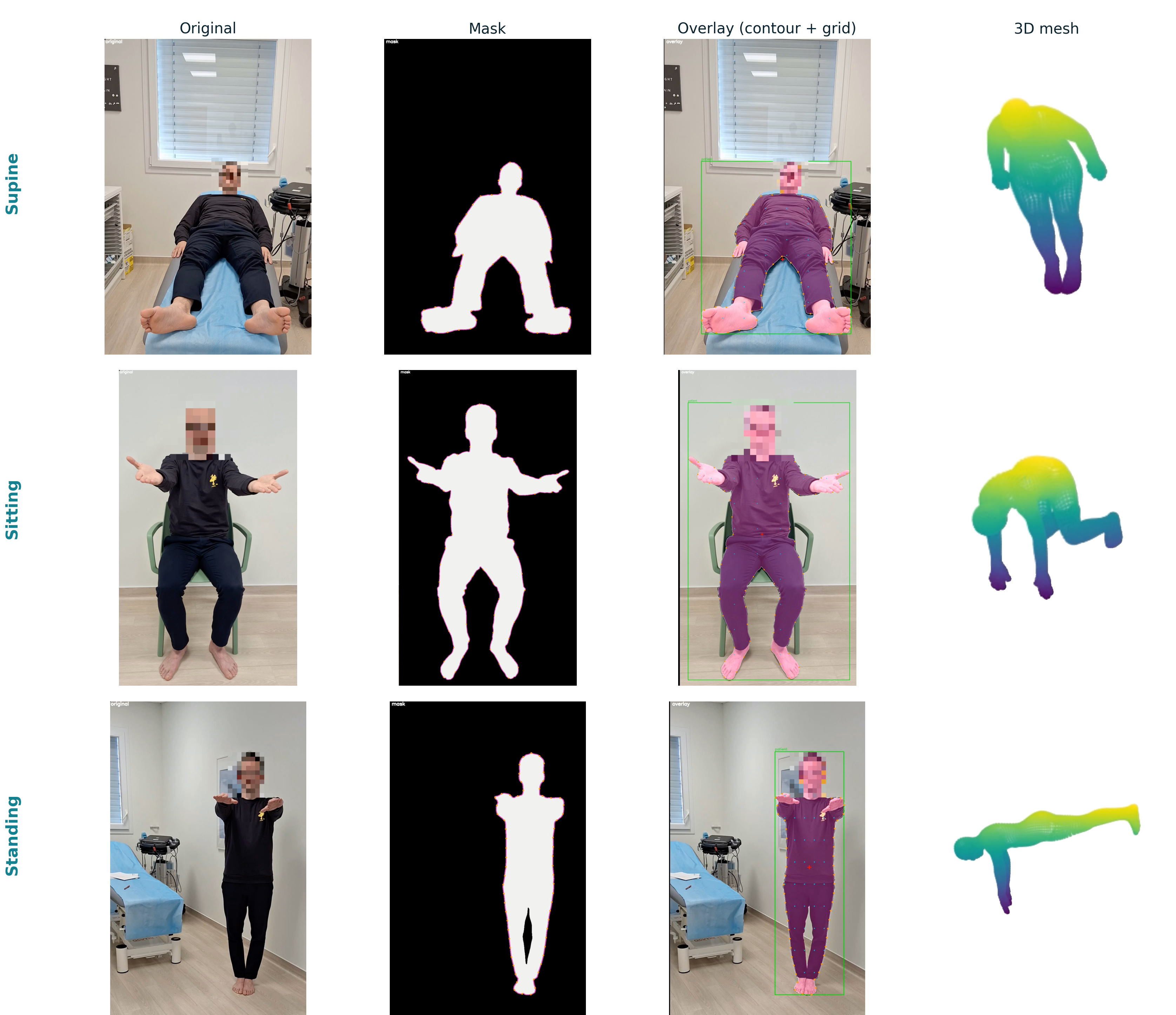}
\caption{\textbf{Dense markerless representation of the patient across examination configurations.} Each row corresponds to one examination condition (supine, sitting, standing). From left to right: the original video frame; the binary patient mask produced by SAM~3 from the open-vocabulary text prompt "person"; the dense overlay used for feature extraction, showing the 64-point resampled body contour and the 12$\times$8 interior grid, with the green box marking the retained patient track; and the corresponding single-image 3D body mesh recovered with SAM~3D Body, in which colour encodes mesh depth. The 3D reconstruction is used for qualitative visualization only and is not part of the quantitative classification pipeline. Faces are pixelated for anonymization.}
\label{fig:2}
\end{figure}

From each patient mask we derived three complementary families of per-frame signals. (i) Fourteen geometric descriptors of the silhouette: centroid coordinates, pixel area, perimeter, bounding-box position and size, aspect ratio, solidity (area divided by convex-hull area), extent (area divided by bounding-box area), principal-axis orientation, and major and minor principal-axis lengths from a principal-component analysis of the mask pixels. (ii) A contour representation: the external silhouette contour resampled to 64 points evenly spaced by arc length, anchored at the topmost point above the centroid (approximately the head) so that point indices correspond consistently across frames, giving 128 contour coordinates. (iii) An interior grid: a 12-by-8 grid of points spanning the patient bounding box, each identified by its normalized position so that it retains the same semantic body location across frames, giving up to 96 interior points (192 coordinates). Points falling outside the mask were recorded as missing. This produces up to 334 raw per-frame signals (Figure~\ref{fig:2}). To complement the 2D silhouette representation and to support interpretation, we additionally reconstructed a full 3D body mesh from single frames using SAM~3D Body (3DB; Meta, released November 2025), a promptable model for single-image full-body human mesh recovery based on the Momentum Human Rig parametric representation~\cite{ref44}. The SAM~3 patient mask was used as the prompt, so that the 3D reconstruction was locked onto the patient. Meshes were reconstructed on sampled keyframes (by default one per second) and rendered as a rotating animation. The 3D reconstruction was used here for qualitative visualization and as a basis for future 3D feature extraction; the quantitative classification results reported below are based on the 2D dense representation (Figure~\ref{fig:2}).

\subsection{Windowing and kinematic descriptors}

Per-frame signals were segmented into sliding temporal windows of 300 frames (10 seconds at 30 frames per second) with a stride of 150 frames (50\% overlap). Within each window, each signal was summarized by 19 kinematic descriptors capturing distributional, temporal, spectral and complexity properties: mean, standard deviation, minimum, maximum, median, range, skewness, kurtosis, signal energy, linear-trend slope, interquartile range, histogram-based entropy, variance, dominant spectral peak frequency, dominant spectral peak amplitude, count of zero-crossings of the first derivative, mean absolute acceleration, Higuchi fractal dimension and permutation entropy. Each signal was scaled using a robust median-and-interquartile-range normalization applied within each 10-second window, before the descriptors were computed, no statistic being estimated on the training set and reapplied at inference. Normalizing within the window removes absolute amplitude, so for tremor the descriptors encode frequency and regularity rather than magnitude.

\subsection{Feature tiers and tier selection}

Because the number of raw signals is large relative to the dataset size, we defined three nested feature schemes (``tiers'') of increasing capacity and selected the operating tier by internal cross-validation rather than fixing it \emph{a priori}.

\begin{itemize}
  \item \textbf{Tier~1 (minimal, $\sim$26 signals):} the 14 geometric descriptors plus 12 posture-derived signals computed from the silhouette contour. The signal count of a tier is not a tuned parameter: it follows mechanically from which of the three families of per-frame measurements (geometric, contour, interior grid) the tier includes and from how they are aggregated, as specified for each tier.
  \item \textbf{Tier~2 (recommended, 81 signals):} Tier~1 augmented by anatomical regional aggregations. The contour is partitioned into seven anatomical zones (cranial, right and left shoulder, right and left upper limb, right and left lower limb) and the interior grid into four regions (upper, middle, lower left and lower right); for each of these eleven regions we compute five aggregate signals: the two coordinates of the regional centroid, radial dispersion, kinetic energy and dominant angle. With the 19 descriptors applied per signal, this gives 11~$\times$~5 = 55 regional signals which, with the 26 Tier-1 signals, make 81 signals and 81~$\times$~19 = 1,539 kinematic features per window. Seven binary indicators of the annotation source and of the examination condition (rater-specific or consensus annotation set; rest, posture or action task; short condition-specific recording) are appended to this block, so the classifier receives 1,546 columns per window.
  \item \textbf{Tier~3 (exhaustive, $\sim$334 signals):} all raw signals (14 geometric, 128 contour and 192 grid coordinates). Maximum information, but the highest overfitting risk on small datasets.
\end{itemize}

The three tiers were compared by patient-grouped cross-validation on the training dataset, and the operating tier was selected on that basis alone, before any evaluation dataset was analysed. The comparison itself and the resulting choice of Tier~2 are reported in the Results.

\subsection{Classifier: a tabular foundation model}

The windowed feature matrix was classified with TabICLv2, a tabular foundation model performing in-context learning for tabular classification~\cite{ref42}. TabICLv2 builds on the TabICL~\cite{ref41} and TabPFN~\cite{ref40} line of work, in which a transformer pre-trained on large numbers of synthetic tabular tasks predicts labels for a new dataset directly from a context of labelled examples, without per-dataset gradient training~\cite{ref40,ref41,ref42}. The different phenomenologies were modelled as eight independent binary one-versus-rest tasks. The in-context set was drawn from the training windows only: for each phenomenology, every window annotated present plus a stratified sample of at most twenty times as many annotated absent, uncertain windows excluded. Of the 5,618 training windows (25 participants, 111 to 313 each), this left 5,201 to 5,307 windows of context for the six more prevalent phenomenologies and 1,911, 483 and 441 for stereotypies, tics and ballismus, in every case well within the context capacity of the model. The model produced per-window probabilities, which were then aggregated to the patient level.

\subsection{Site-level calibration of the patient-level decision}

When the framework is deployed at a new site, the trained model itself remains untouched: only the last step: how per window scores are turned into a single present/absent call for each patient, is re-tuned using a handful of patients from that site. When transferring the backbone to an evaluation dataset, only the final patient-level decision step was adapted locally. This step aggregates per-window probabilities into a single patient-level probability per phenomenology and applies a decision threshold. Calibration consisted of choosing, per phenomenology, the window-to-patient aggregation operator (among high-percentile and maximum operators) and the decision threshold, using only a small set of calibration patients from that dataset (five of the twelve patients in the pediatric dataset\_2, five of the twenty in the adult dataset\_3, and one of the two in dataset\_1). The backbone and the feature representation were never modified.

The calibration objective followed the multi-label, patient-level criterion established in the previously published pipeline~\cite{ref36,ref37}. For a candidate configuration producing predicted label sets against the clinician reference, the score combined a multi-label agreement term with two penalties:

\begin{equation}
\mathrm{score} = J(Y,\hat{Y}) - \lambda_{\mathrm{FP}}\cdot \mathrm{FP}_{\mathrm{excess}} - \lambda_{\mathrm{FN}}\cdot \mathrm{FN}_{\mathrm{key}}
\label{eq:score}
\end{equation}

\noindent where $Y$ and $\hat{Y}$ are the true and the predicted label sets of a patient, $J$ is a multi-label agreement measure (Jaccard index by default, with exact-match and macro-F1 available), $\mathrm{FP}_{\mathrm{excess}}$ penalizes predicting more positive labels than truly present (normalized by the eight phenomenologies), and $\mathrm{FN}_{\mathrm{key}}$ penalizes missed positives on a set of clinically critical phenomenologies (dystonia, myoclonus, chorea). The penalty weights were $\lambda_{\mathrm{FP}}$ = 0.35 and $\lambda_{\mathrm{FN}}$ = 0.35. In plain terms, this score rises when the predicted phenomenologies match the clinician's labels and it is reduced whenever the model calls extra phenomenologies that are not present or misses one of the three phenomenologies judged most clinically important. The configuration was optimized by coordinate descent ( a systematic search that adjusts one setting at a time, keeping the others fixed, and repeats until the score stops improving) over aggregation operators (high percentiles and maximum) and per-phenomenology thresholds, with clinically motivated threshold guard-rails: minimum thresholds for phenomenologies that are low-prevalence in the training set and easily over-called (for example tremor and tics 0.35, ballismus 0.30, stereotypies 0.25, athetosis 0.20) and maximum thresholds for the frequently dominant phenomenologies (dystonia 0.55, myoclonus 0.50, chorea 0.55). These guard-rails prevent the optimizer from exploiting degenerate all-positive or all-negative rules on small calibration sets. The calibration patients were selected by the rating clinicians a priori, before any model prediction or held-out result was examined, and the selection was then frozen. Three explicit criteria were applied, in this order. First, phenotypic coverage: the subset had to contain, collectively, at least one positive example of each phenomenology present in that dataset, so that every calibrated threshold could be informed by affected patients. Second, annotation reliability: preference was given to patients whose phenomenology was scored concordantly by the raters, so that thresholds were anchored on clinically unambiguous cases rather than on borderline ones. Third, recording quality: patients whose videos allowed the whole body to be seen throughout the examination were preferred, since incomplete framing degrades the silhouette representation. Patient identity, age, etiology, stimulation status and model output played no part in the selection.

\subsection{Exhaustive exploration of calibration subsets}

To characterize how strongly the choice of calibration patients affects performance, and to test the robustness of the approach, we additionally evaluated calibration subsets exhaustively (or, where the number of combinations was large, by extensive sampling). For each candidate calibration subset, we re-ran the calibration objective and recorded held-out performance, building a full distribution of achievable performance rather than relying on a single subset. To avoid circularity, the headline results use the a priori clinician-selected sets.

\subsection{Metrics and statistical analysis}

Performance was assessed at the patient level under the multi-label formulation. We report the Hamming accuracy (the fraction of the eight phenomenology labels predicted correctly, averaged over labels and patients) and the Jaccard index (intersection over union of predicted and true positive labels). Both are computed per patient and then averaged over patients. We additionally report true-positive, true-negative, false-positive and false-negative counts pooled over patients and labels, that is summed rather than averaged, both overall and per phenomenology, so that the behavior of each phenomenology is visible rather than hidden in an average. Results are reported separately for the held-out patients (primary external validity), for all patients, and, where relevant, for the calibration patients. For the exhaustive subset analysis, we report medians and interquartile ranges of the held-out Jaccard index across subsets. Uncertainty was quantified by bootstrap resampling of the held-out patients (2,000 resamples, fixed seed), the 2.5th and 97.5th percentiles being reported as a 95\% confidence interval. Because the Jaccard index has no prevalence-independent chance level, its expectation depending on label prevalence and on the decision rule, it is benchmarked against an empirical null obtained by permuting the patient-level reference label sets across patients with the predicted label sets held fixed (2,000 permutations), of which we report the median and the 95th percentile.

\subsection{Hardware and software}

All video processing and model inference were run on a single workstation (Windows; NVIDIA GeForce RTX 5080 GPU, 16 GB VRAM, Blackwell architecture) using Meta SAM~3 (sam3.pt, approximately 3.45 GB) and SAM~3D Body (facebook/sam-3d-body-dinov3) under PyTorch (nightly build with CUDA 12.8, required for Blackwell sm\_120 support). Feature extraction, classification with TabICLv2 and calibration were implemented in Python 3.11.

\section{Results}

\subsection{Dense markerless representation and 3D recovery versus sparse skeleton representation}

SAM~3 produced stable, dense patient silhouettes across lying, sitting and standing postures, and the automatic patient track excluded the clinician when present (Figure~\ref{fig:2}). The resampled contour and interior grid provided regional motion information across the whole-body surface, including the trunk and limb outlines that a sparse skeleton does not capture. Single-image 3D body-mesh recovery with SAM~3D Body produced anatomically plausible meshes from the same frames, prompted by the SAM~3 mask, confirming that a 3D representation can be obtained from the same routine monocular video for future quantitative use.

\subsection{Internal cross-validation and tier selection}

\begin{figure}[tbp]
\centering
\includegraphics[width=\textwidth]{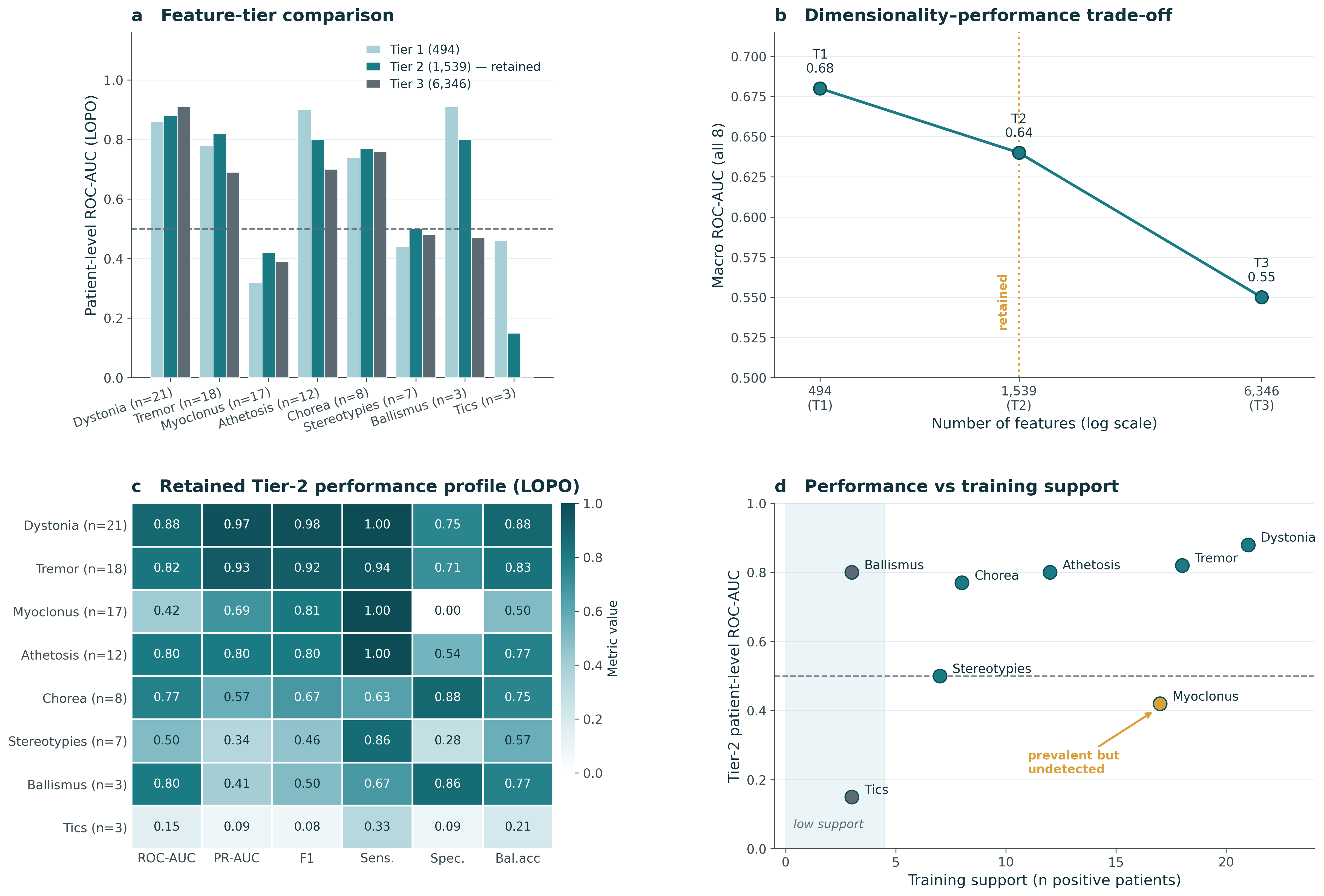}
\caption{\textbf{Internal cross-validation and feature-tier selection on the training dataset.} Performance was estimated by patient-grouped leave-one-patient-out cross-validation (25 folds), with the out-of-fold predictions of all folds assembled into a single prediction set before computing patient-level metrics; n indicates the training support, that is the number of patients in whom the phenomenology was annotated present by either rater, a less strict rule than the two-rater prevalences given in the Methods. (a) Patient-level ROC-AUC per phenomenology for the three nested feature tiers (494, 1,539 and 6,346 kinematic features); the dashed line marks chance (0.5). (b) Macro-averaged ROC-AUC across all eight phenomenologies as a function of feature-space dimensionality; performance declines as features are added, indicating overfitting in the highest-dimensional tier, and Tier~2 (1,539 features) was retained as the operating configuration. (c) Full performance profile of the retained Tier~2 configuration (ROC-AUC, PR-AUC, F1, sensitivity, specificity and balanced accuracy) for each phenomenology. (d) Relationship between training support and Tier~2 ROC-AUC: performance increased with the number of positive patients, with the notable exception of myoclonus, which was prevalent yet undetected, reflecting a representational rather than a statistical limitation; the lowest-support phenomenologies (ballismus and tics, three positive patients each) lie in the shaded low-support regime and are not interpreted.}
\label{fig:3}
\end{figure}

Internal validity and the operating configuration were established by patient-grouped leave-one-patient-out cross-validation on the 25-participant training dataset (25 folds; out-of-fold predictions pooled before computing patient-level metrics). We used this to select among the three nested feature tiers as described in the method section. (Figure~\ref{fig:3}a, b; Supplementary Table~\ref{tab:S1}). Macro-averaged ROC-AUC across all eight HMDs was highest for the smallest tier (Tier~1, 0.68; Tier~2, 0.64; Tier~3, 0.55), but this ranking was driven by the rarest labels, estimated on three to seven positive patients. Restricted to the five forms of phenomenology with adequate support (dystonia, tremor, myoclonus, chorea, athetosis; 8 to 21 positive patients), Tier~2 showed the best macro ROC-AUC (0.74, versus 0.72 for Tier~1 and 0.69 for Tier~3) and was retained: it describes movement separately in each body region, which the smallest tier cannot do, while limiting the overfitting evident in the 6,346-feature Tier~3 (Figure~\ref{fig:3}b). Under Tier~2, the dense representation separated the two dominant phenomenologies most reliably at the patient level (dystonia ROC-AUC 0.88, PR-AUC 0.97; tremor 0.82 and 0.93) and recovered chorea and athetosis at a clinically useful level (ROC-AUC 0.77 and 0.80; Figure~\ref{fig:3}c). At the tuned operating points, sensitivity reached 1.00 for dystonia, athetosis and myoclonus and 0.94 for tremor, while specificity varied with prevalence (Figure~\ref{fig:3}d).~Myoclonus, although the third most represented phenomenology (17 positive patients), was the principal failure mode (ROC-AUC 0.42): its per-window probabilities did not separate affected from unaffected patients, and it was called present in every patient (sensitivity 1.00, specificity 0.00).

\subsection{External transfer with clinician-selected calibration sets}

Using the \emph{a priori} clinician-selected calibration sets, the shared backbone transferred to both external datasets with recalibration of the decision step alone. Because the number of raters differed across datasets, each result is reported across a ladder of options as previously defined, from the most permissive to rater consensus; we treat this ladder as a sensitivity analysis and the consensus-based (restricted) options, which remove ambiguous calls, as the primary reference.

\begin{figure}[tbp]
\centering
\includegraphics[width=\textwidth]{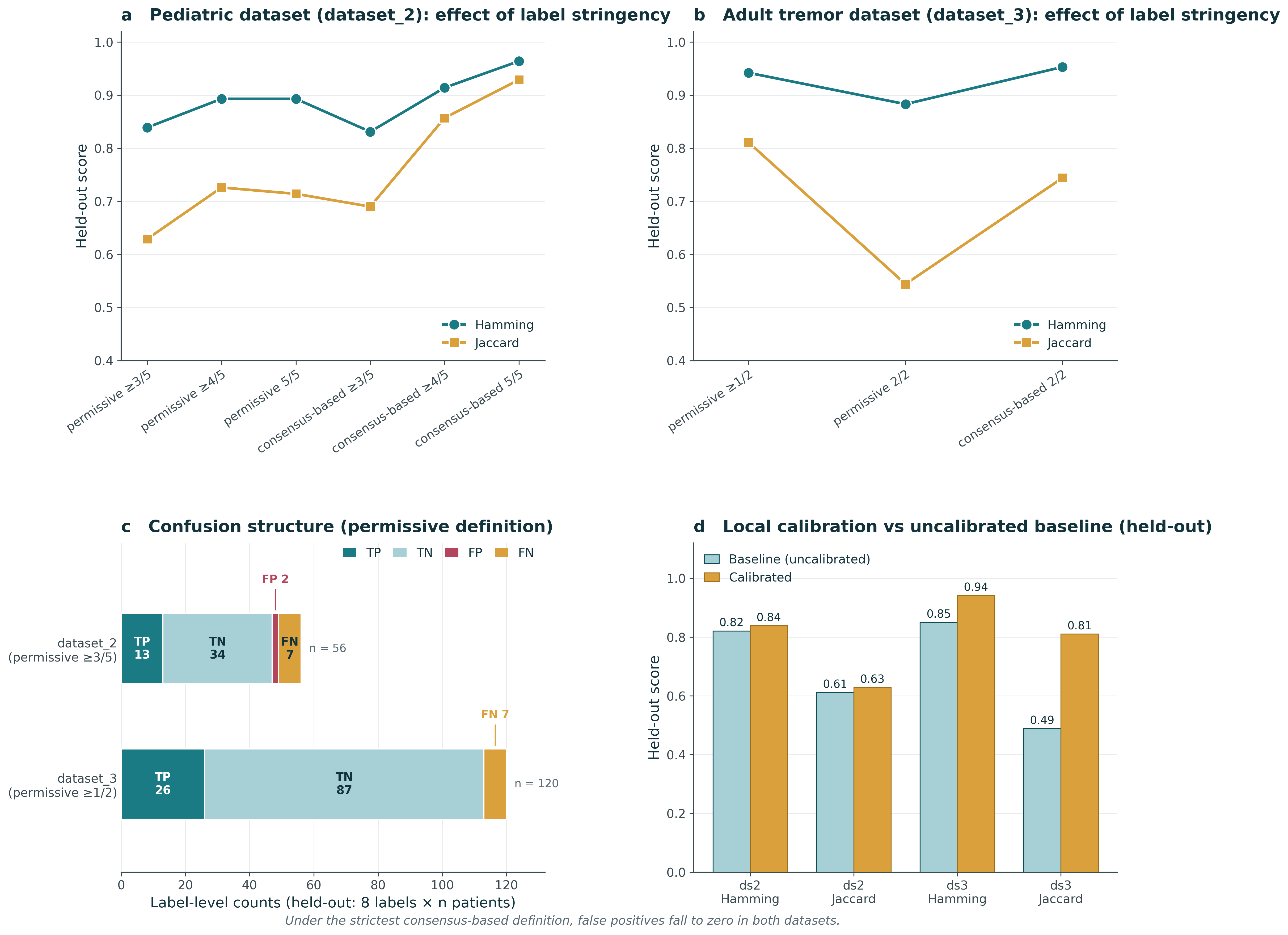}
\caption{\textbf{External validation with clinician-selected calibration sets.} The shared frozen backbone was transferred to the evaluation datasets by recalibrating only the decision step on an a priori, clinician-selected calibration set, then evaluated on the remaining held-out patients. (a, b) Held-out Hamming accuracy and Jaccard index as the label definition is tightened from the most permissive (at least three of five raters in a; at least one of two in b) toward rater consensus, for the pediatric (a) and adult (b) datasets; performance increases as labelling approaches consensus. (c) Label-level confusion structure under the permissive definition for both datasets (held-out; 8 phenomenologies $\times$ n patients); under the strictest consensus-based definition, false positives fall to zero in both datasets (Tables~\ref{tab:2}, \ref{tab:3}). (d) Held-out performance of the locally calibrated decision rule versus the uncalibrated generic baseline on the same held-out patients (permissive definition): calibration yields a large gain on the adult dataset\_3 and a modest gain on the pediatric dataset\_2.}
\label{fig:4}
\end{figure}

\begin{figure}[tbp]
\centering
\includegraphics[width=0.92\textwidth]{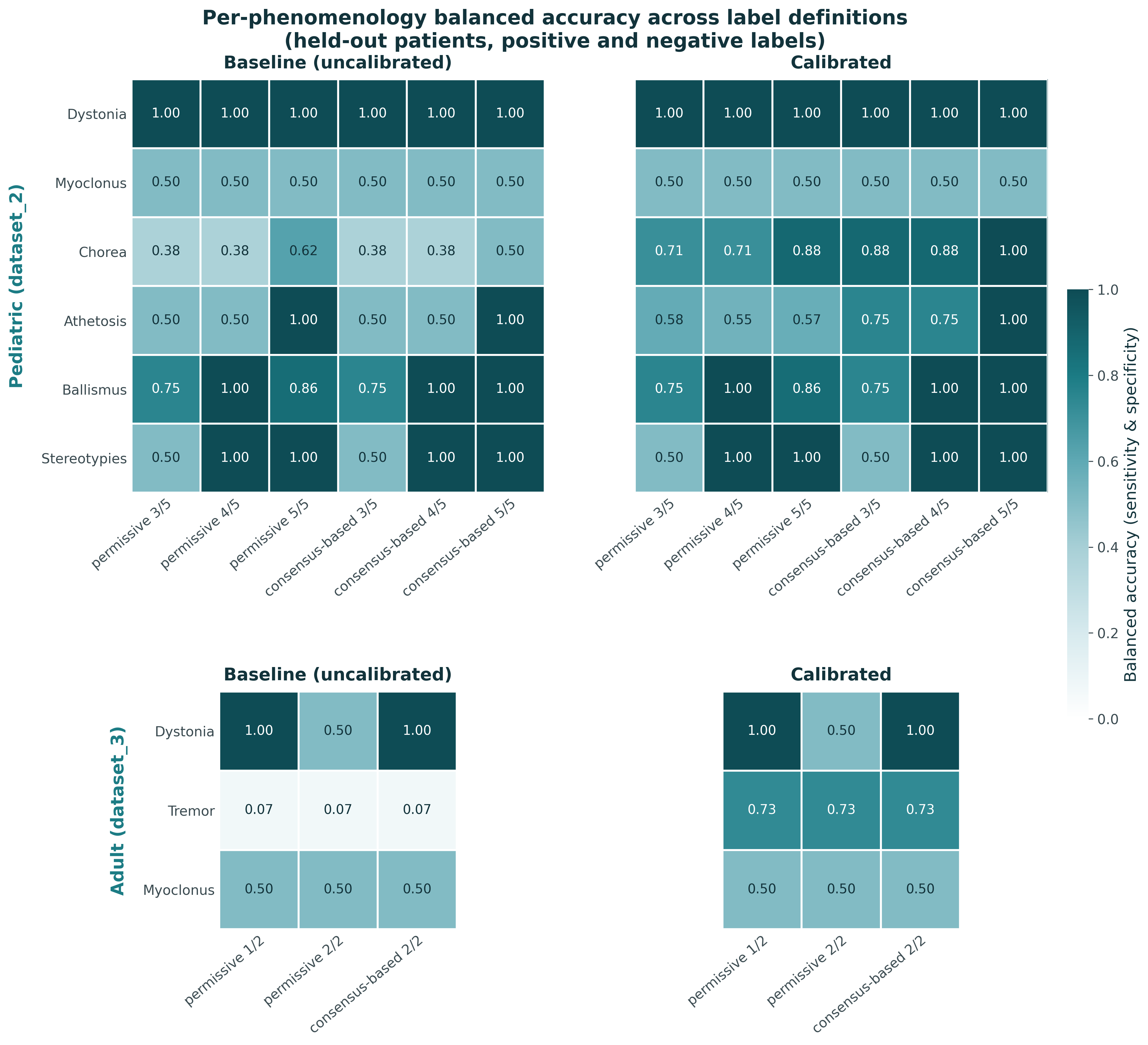}
\caption{\textbf{Per-phenomenology balanced accuracy across label definitions, before and after local calibration.} Balanced accuracy (the mean of sensitivity and specificity, weighting equally the correct detection of present phenomenologies and the correct rejection of absent ones) for each phenomenology and each label definition, on the held-out patients, for the uncalibrated baseline (left) and the locally calibrated decision rule (right), in the pediatric (top) and adult (bottom) datasets. Local calibration improves the recovery of chorea and athetosis in the pediatric dataset\_2 and, most markedly, tremor in the adult dataset\_3 (0.07 to 0.73), whereas dystonia is classified near-perfectly throughout and myoclonus remains at chance level (0.50) across datasets and definitions. Only phenomenologies with positive held-out patients are shown.}
\label{fig:5}
\end{figure}

For dataset\_2, on the seven held-out pediatric patients, performance increased monotonically within the consensus-based family as the definition was tightened toward consensus, from a Hamming accuracy of 0.84 (95\% CI 0.75 to 0.93) and a Jaccard index of 0.63 (0.45 to 0.83) under the most permissive option to 0.96 (0.89 to 1.00) and 0.93 (0.79 to 1.00) under the consensus-based one, with no false positives and a single false negative (Figure~\ref{fig:4}a, Table~\ref{tab:2}). Under the consensus-based option, dystonia was classified perfectly (7/7) and chorea was fully recovered (3/3) with no false positives on any label; the only error was one missed myoclonus call. Under the most permissive option, dystonia remained perfectly classified with myoclonus the main weakness. For dataset\_3, on the 15 held-out adults, both prevalent phenomenologies were recovered: under the at-least-one-rater option the model reached a Hamming accuracy of 0.94 (95\% CI 0.90 to 0.98) and a Jaccard index of 0.81 (0.69 to 0.92) with no false positives (Figure~\ref{fig:4}b, Table~\ref{tab:3}). Dystonia was detected in 15/15 held-out patients and tremor was well detected (11 true positives, 4 false negatives), a phenomenology that had been effectively non-evaluable in the pediatric dataset (Figure~\ref{fig:5}). Myoclonus was again the principal failure mode (no true positives, three false negatives), and the phenomenologies essentially absent from this cohort were correctly predicted as negative. Two-rater permissive option introduced false positives, removed under the consensus-based option, restoring a Hamming accuracy of 0.95 (0.92 to 0.98). Across both datasets, the error profile was conservative, with false positives rare and eliminated entirely under the consensus-based options (Figure~\ref{fig:4}c). Local recalibration improved held-out performance over the uncalibrated generic rule evaluated on the same patients, substantially on the adult sample (Hamming 0.85 to 0.94; Jaccard 0.49 to 0.81) and modestly on the pediatric sample (Hamming 0.82 to 0.84; Jaccard 0.61 to 0.63; Figure~\ref{fig:4}d), consistent with the adult dataset benefiting most from site-specific threshold adaptation.

\begin{table}[tbp]
\centering
\caption{\textbf{Held-out multi-label performance on the pediatric dataset\_2 (n=7) under permissive and consensus-based label definitions.} Performance increases as the labelling is tightened toward rater consensus; under all consensus-based definitions there are no false positives. TP/TN/FP/FN are label-level counts over the 8 phenomenologies and 7 held-out patients.}
\label{tab:2}
\small
\begin{tabular}{lcccccc}
\toprule
\textbf{Label definition} & \textbf{Hamming} & \textbf{Jaccard} & \textbf{TP} & \textbf{TN} & \textbf{FP} & \textbf{FN} \\
\midrule
Permissive $\geq$3/5 & 0.839 & 0.629 & 13 & 34 & 2 & 7 \\
Permissive $\geq$4/5 & 0.893 & 0.726 & 12 & 38 & 3 & 3 \\
Permissive 5/5 & 0.893 & 0.714 & 10 & 40 & 5 & 1 \\
Consensus-based $\geq$3/5 & 0.831 & 0.690 & 13 & 22 & 0 & 7 \\
Consensus-based $\geq$4/5 & 0.914 & 0.857 & 12 & 22 & 0 & 3 \\
Consensus-based 5/5 & 0.964 & 0.929 & 10 & 22 & 0 & 1 \\
\bottomrule
\end{tabular}
\end{table}

\begin{table}[tbp]
\centering
\caption{\textbf{Held-out multi-label performance on the adult tremor-prominent dataset\_3 (n=15) under rater-adaptive label definitions.} Tremor is well detected in this dataset, in contrast to the pediatric dataset\_2 where it was rare. The consensus (2/2) permissive  definition introduces false positives that are removed under the consensus-based  definition. TP/TN/FP/FN are label-level counts over the 8 phenomenologies and 15 held-out patients.}
\label{tab:3}
\small
\begin{tabular}{lcccccc}
\toprule
\textbf{Label definition} & \textbf{Hamming} & \textbf{Jaccard} & \textbf{TP} & \textbf{TN} & \textbf{FP} & \textbf{FN} \\
\midrule
Permissive $\geq$1/2 & 0.942 & 0.811 & 26 & 87 & 0 & 7 \\
main 2/2 (consensus) & 0.883 & 0.544 & 17 & 89 & 9 & 5 \\
Consensus 2/2 & 0.953 & 0.744 & 17 & 87 & 0 & 5 \\
\bottomrule
\end{tabular}
\end{table}

\subsection{Robustness across all calibration subsets}

\begin{figure}[tbp]
\centering
\includegraphics[width=\textwidth]{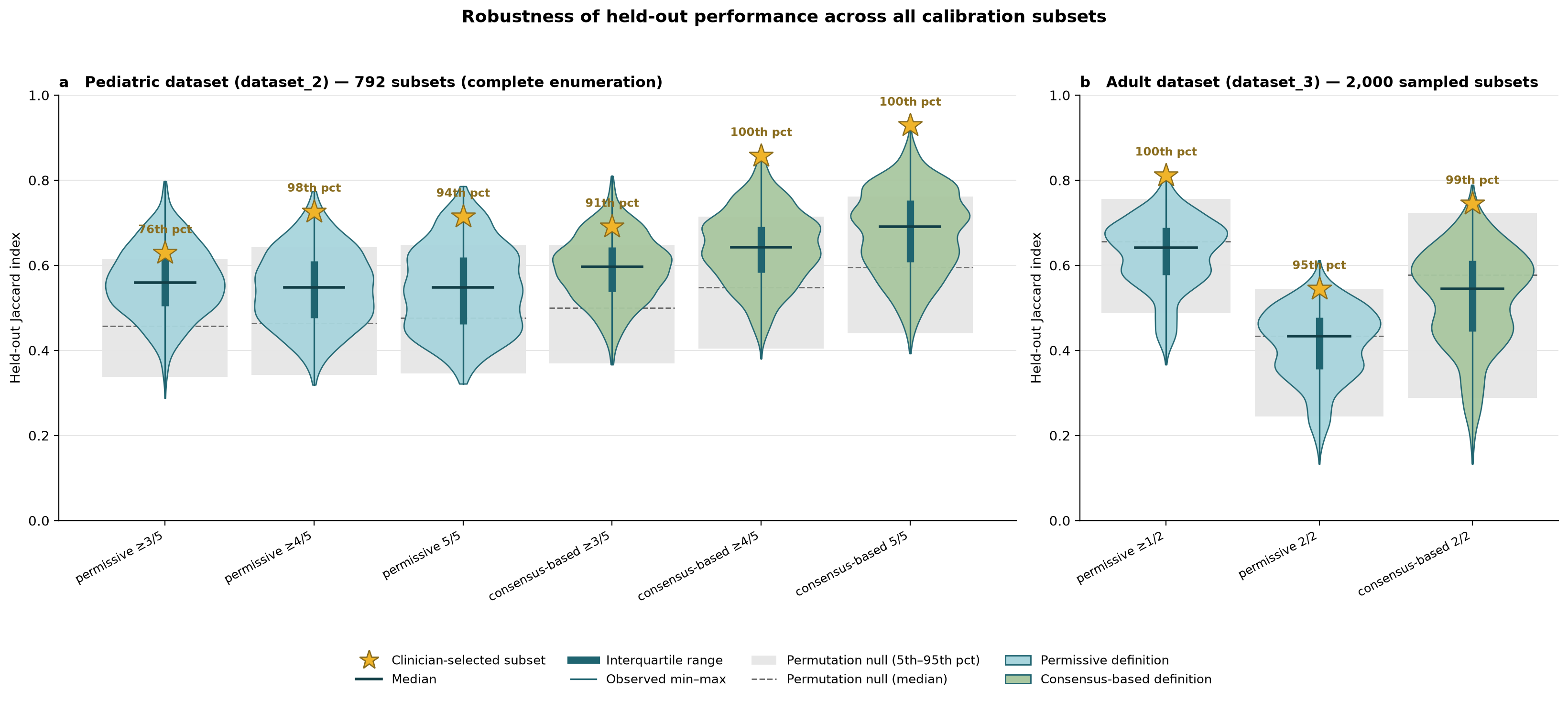}
\caption{\textbf{Robustness of held-out performance across all calibration subsets.} Distribution of the held-out Jaccard index over all evaluated calibration subsets (pediatric dataset\_2, 792 subsets by complete enumeration; adult dataset\_3, 2,000 sampled subsets), for each label definition. Violins show the full distribution with the median; the gold star marks the a priori clinician-selected subset used for the headline results; the grey band and its dashed line are the 5th to 95th percentile range and the median of the label-permutation null (200 permutations per calibration subset, pooled over all subsets), which replaces the fixed 0.5 line because the Jaccard index has no prevalence-independent chance level. The pediatric distributions sit above their null, the adult ones overlap theirs. Performance is stable across the space of calibration choices, the clinician-selected subset sits in the upper part of each distribution, and tightening the definition toward consensus shifts the distribution upward in the pediatric dataset\_2, whereas full two-rater consensus in the adult dataset\_3 (2/2, permissive) is more demanding and widens the lower tail.}
\label{fig:6}
\end{figure}

To test whether the headline results depended on a fortunate choice of calibration patients, we evaluated calibration subsets exhaustively (dataset\_2: 792 subsets, complete enumeration of all five-patient combinations; dataset\_3: 2,000 sampled five-patient subsets), recording the held-out Jaccard index for each (Table~\ref{tab:4}, Figure~\ref{fig:6}).

\begin{table}[tbp]
\centering
\caption{\textbf{Distribution of the held-out Jaccard index across all evaluated calibration subsets (dataset\_2 fully enumerated, 792 five-patient subsets; dataset\_3 sampled, 2,000 five-patient subsets).} Dispersion across the evaluated subsets is reported as the interquartile range (Q1--Q3) and the 5th--95th percentile range; each value is computed over the 792 enumerated (dataset\_2) or 2,000 sampled (dataset\_3) subsets, one held-out Jaccard index per subset. The a priori clinician-selected subset used for the headline results is shown with its percentile rank within each distribution. It falls in the upper part of every distribution, which is consistent with informed clinical selection, fixed before any held-out evaluation, outperforms the typical subset.}
\label{tab:4}
\footnotesize
\begin{tabular}{lccccc}
\toprule
\textbf{Cohort / definition} & \textbf{Subsets (n)} & \textbf{Median Jaccard} & \textbf{IQR (Q1--Q3)} & \textbf{5th--95th pct.} & \textbf{Clinician subset (pct.)} \\
\midrule
dataset\_2, permissive $\geq$3/5 & 792 & 0.560 & 0.50--0.62 & 0.43--0.69 & 0.629 (76th) \\
dataset\_2, consensus 5/5 & 792 & 0.690 & 0.61--0.75 & 0.51--0.81 & 0.929 (100th) \\
dataset\_3, permissive $\geq$1/2 & 2,000 & 0.642 & 0.58--0.69 & 0.46--0.74 & 0.811 (100th) \\
dataset\_3, permissive 2/2 & 2,000 & 0.433 & 0.36--0.48 & 0.24--0.53 & 0.544 (95th) \\
\bottomrule
\end{tabular}
\end{table}

Two conclusions emerge. First, the approach does not rely on a single hand-picked subset: across hundreds to thousands of calibration choices, the median held-out Jaccard varied little across datasets and definitions (0.56 to 0.69 in the pediatric sample; 0.43 to 0.64 in the adult sample), and the bulk of each distribution stayed in a clinically useful range. Against the label-permutation null, however, the two datasets differ. In the pediatric dataset\_2 the observed median exceeded the null median at every definition (0.56 against 0.46 at the most permissive level, 0.69 against 0.60 at unanimity), whereas in the adult dataset\_3 it was at or below it (0.64 against 0.66, 0.43 against 0.43, 0.54 against 0.58), so the held-out Jaccard index in that cohort is not distinguishable from chance. Dystonia and tremor are annotated in almost every adult patient, which leaves a multi-label overlap index very little room to demonstrate patient-level discrimination. Second, and equally important, the choice of calibration patients is not neutral: the distributions are dispersed (5th-95th percentile spans of roughly 0.25 in width), so a poorly chosen calibration set can fall near the lower tail while a well-chosen one reaches the top. The \emph{a priori} clinician-selected subsets used for the headline results sat consistently in the upper part of these distributions (76th, 100th, 100th and 95th percentile across the four dataset-definition combinations).

\subsection{What the model attends to, by phenomenology}

To understand why the approach works and where it fails, we compared positive and negative windows in the Tier-2 training data (effect size, Cohen's d) and mapped these signals onto body regions (Table~\ref{tab:5}).

\begin{table}[tbp]
\centering
\caption{\textbf{Most discriminative kinematic signals per phenomenology, with anatomical localization and clinical interpretation.} Discriminative power is the effect size (|Cohen's d|) separating positive from negative windows in the Tier-2 training data; the strongest signal per phenomenology is reported. The region named is that of the single strongest signal, which need not be the region with the highest regional maximum in Figure~\ref{fig:7}: for tremor the strongest signal is axial (0.29) whereas the highest regional maximum is the interior grid (0.30). Effect sizes are modest in absolute terms because each signal is one of many, but their anatomical pattern is phenomenology-specific and clinically coherent.}
\label{tab:5}
\footnotesize
\renewcommand{\arraystretch}{1.25}
\begin{tabularx}{\textwidth}{@{}>{\raggedright\arraybackslash}p{2.1cm}>{\raggedright\arraybackslash}p{2.3cm}>{\raggedright\arraybackslash}p{3.8cm}>{\raggedright\arraybackslash}X@{}}
\toprule
\textbf{Phenomenology} & \textbf{Body region(s)} & \textbf{Representative signals ($|$Cohen's $d|$)} & \textbf{Clinical interpretation} \\
\midrule
\textbf{Chorea} & Head, proximal arms & cranial-region displacement (0.59), cranial-segment movement amplitude (0.54) & Irregular, flowing displacement of the head and proximal upper limbs. \\
\textbf{Athetosis} & Lower interior body, shoulders & lower interior-grid radial dispersion (0.60), shoulder asymmetry (0.55) & Slow distal writhing captured by dispersion of interior body points. \\
\textbf{Stereotypies} & Interior body & interior-grid centroid / dominant angle (0.65) & Patterned, repetitive interior motion. \\
\textbf{Myoclonus} & Whole-body (diffuse) & major axis (0.44), silhouette area (0.42) & Abrupt global silhouette changes; weakest and most diffuse signature, consistent with poor detectability of brief jerks. \\
\textbf{Tremor} & Trunk / axial & body-axis orientation (0.29), principal-axis angle (0.29) & Oscillation of the principal body axis. \\
\textbf{Dystonia} & Trunk / global posture & vertical extent (0.30), bounding-box height (0.30) & Sustained changes in overall posture and silhouette height. \\
\bottomrule
\end{tabularx}
\end{table}

\begin{figure}[tbp]
\centering
\includegraphics[width=0.85\textwidth]{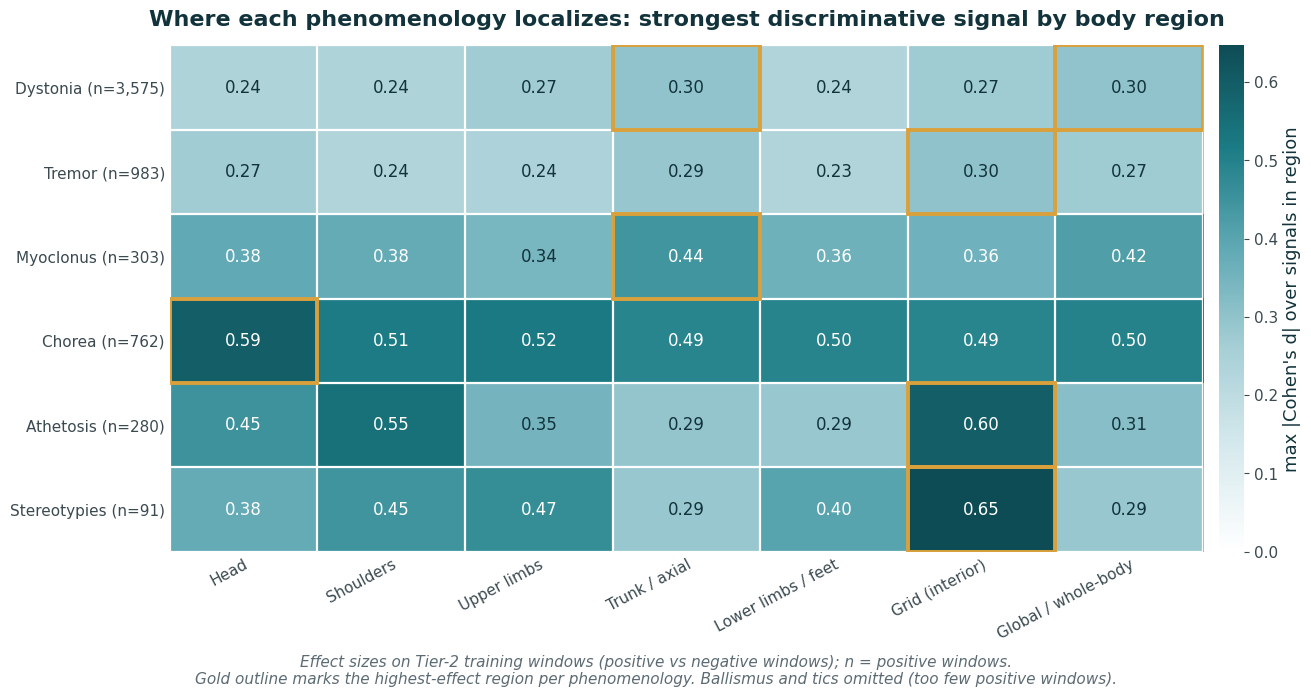}
\caption{\textbf{Where each phenomenology localizes: strongest discriminative signal by body region.} Maximum absolute effect size (|Cohen's d|) separating windows annotated present from windows annotated absent, among the Tier-2 signals assigned to each region, on the training dataset; n is the number of positive windows. The seven contour zones and four grid regions are pooled into head, shoulders, upper limbs, trunk and axial, lower limbs and feet, interior grid, and whole-body signals belonging to no single zone. The gold outline marks each phenomenology's highest-effect region. Ballismus and tics are omitted, having too few positive windows to be interpretable.}
\label{fig:7}
\end{figure}

The discriminative signals, that is the kinematic measurements whose values differ most between windows in which a phenomenology was annotated present and windows in which it was annotated absent, were phenomenology-specific and clinically coherent. Chorea localized to the cranial segment and proximal upper limbs (cranial segment displacement and cranial-segment movement amplitude, $|d|$ up to 0.59), athetosis to lower-body interior-point dispersion and shoulder asymmetry ($|d|$ up to 0.60), stereotypies to patterned interior-grid motion ($|d|$ up to 0.65), tremor to oscillation of the principal body axis, that is to the trunk and axial zone ($|d|$ $\approx$ 0.29), and dystonia to sustained changes in overall posture and silhouette height ($|d|$ $\approx$ 0.30). Myoclonus produced the most diffuse signature, with no region reaching a strong effect size ($|d|$ $\leq$ 0.44).

We also asked whether effective calibration subsets share a measurable structural signature, relating each subset's phenotypic coverage and mean inter-rater agreement to held-out Jaccard. These properties explained little of the variation (coverage: Spearman $\rho$ = 0.17 in the pediatric data, and non-significant to slightly negative in the adult data; agreement: $\rho$ $\leq$ 0.08, and negative in the adult dataset). Calibration-set selection therefore cannot be reduced to a simple coverage- or agreement-based rule; the practical recommendation remains to follow the \emph{a priori} clinical criteria used here, while recognizing that they are not on their own strong quantitative predictors of held-out performance.

\section{Discussion}

We show that two foundation models, an open-vocabulary segmentation model for dense marker-less body description and a pretrained tabular model for in-context classification, can be combined into a transferable, multi-label framework for the simultaneous phenotyping of co-occurring hyperkinetic movement disorders. A single backbone, trained once on standardized adult examinations and never retrained, transferred to an independent routine-care pediatric cohort and to an independent adult, tremor- and dystonia-dominant dataset, with only the patient-level decision step recalibrated locally. This supports our program's train-once, calibrate-then-deploy strategy. The representation moved from sparse keypoints to a dense whole-body description and the cohorts extended from a pediatric setting to an adult tremor-predominant one.

Three features separate this work from the existing literature. First, its target. Nearly all video-based movement-disorder research addresses PD and quantifies a single sign, most often brady-kinesia through hand pose estimation~\cite{ref2,ref21,ref27}, while the small HMD literature addresses isolated phenomenologies, principally tics and focal dystonia, in dedicated tasks~\cite{ref25,ref30,ref32,ref33}. We address eight hyperkinetic phenomenologies simultaneously and as a multi-label problem, which is how they actually present at the bedside. Second, its acquisition setting. Unlike multi-modal laboratory protocols combining electromyography, accelerometry and three-dimensional video~\cite{ref34}, our input is a single consumer camera, and for one evaluation cohort it is unstandardized routine clinical footage. Third, its adaptation regime. Because both the visual and the classification stages are frozen foundation models, transfer to a new population required recalibration of the decision step alone. That combination is what allows a training set of 21 patients to inform predictions in populations differing in age, etiology and recording conditions, a scale of dataset that conventional supervised training could not exploit.

\subsection{Advance over the sparse-keypoint pose estimation of our previous work, based on YOLOv8-Pose}

Because the pediatric held-out patients are shared with the predecessor transfer study~\cite{ref36,ref37}, which used sparse YOLOv8-Pose key points~\cite{ref45}, the effect of the representation can be isolated. Read along the agreement-based ladder, which holds the label family constant and varies only the required level of rater agreement, the two representations cross over (Table~\ref{tab:6}). At the most permissive level the sparse and dense pipelines were comparable; as agreement tightened, the dense representation improved monotonically while the sparse one peaked at intermediate agreement and then declined, so the dense representation was clearly ahead at unanimity under both label definitions. Dense segmentation therefore does not inflate lenient-operating-point performance; it improves robustness precisely on the high-confidence, consensus labels that clinicians agree on. Adding an adult cohort in which tremor is near-universal also made tremor evaluable for the first time in this program, complementing the pediatric dataset in which tremor was rare~\cite{ref46}.

\begin{table}[tbp]
\centering
\caption{\textbf{Held-out Jaccard index of the dense and sparse representations on the same seven held-out pediatric patients, across label definitions.} The sparse values are those reported for the locally calibrated deployment of the predecessor transfer study. Within the agreement-based family the two representations cross over: the sparse pipeline is marginally ahead at the most permissive level, the dense representation overtakes it at four of five raters, and the gap widens at unanimity. The dense representation increases monotonically with the required level of agreement, whereas the sparse one peaks at four of five and then declines.}
\label{tab:6}
\small
\begin{tabular}{lcc}
\toprule
\textbf{Label definition} & \textbf{Dense (SAM 3, this study)} & \textbf{Sparse keypoints (YOLOv8-Pose)} \\
\midrule
Permissive present/absent, $\geq$3/5 & 0.629 & 0.633 \\
Permissive present/absent, $\geq$4/5 & 0.726 & not reported \\
Permissive present/absent, 5/5 & 0.714 & 0.512 \\
Consensus, $\geq$3/5 & 0.690 & 0.714 \\
Consensus, $\geq$4/5 & 0.857 & 0.786 \\
Consensus, 5/5 & 0.929 & 0.762 \\
\bottomrule
\end{tabular}
\end{table}

\subsection{A two-dimensional representation is a defensible design choice, not merely a limitation}

Whether a three-dimensional representation would improve phenotyping is not a settled question. Three-dimensional analysis, through depth-sensing cameras or by lifting 2D keypoints into 3D, adds value chiefly when the movement is highly articulated or occurs out of the image plane: it is preferred for endpoints requiring absolute joint angles, for tracking segments that change orientation relative to the camera (for example the hands during finger-to-nose testing), and where head or body rotation would otherwise obscure the feature being measured~\cite{ref6,ref47,ref48}. In HMDs, 3D acquisition has been used to build classification tools spanning dystonia, myoclonus and tremor~\cite{ref34}, 3D facial landmarks have enabled head-pose normalization for eye-tic detection~\cite{ref30}, and 3D hand-pose estimation has been applied to finger-tapping and kinetic tremor~\cite{ref16}; for gait, 3D-informed models estimate severity from monocular video and recover joint-angle parameters that 2D pipelines capture less reliably~\cite{ref17}.

These advantages are, however, task-dependent, and the choice of examination task is itself critical for the best visibility of the different phenomenologies~\cite{ref12}, and for the signals most relevant to several of our phenomenologies a 2D representation is competitive or preferable. Current 3D lifting from monocular video is comparatively immature and introduces noise in the depth dimension; because tremor and myoclonus involve subtle, high-frequency, oscillatory phenomena, this depth-related jitter can obscure rather than reveal the signal, and 2D pipelines have been favoured for these features~\cite{ref15}. Benchmarked against optoelectronic motion capture, 2D pipelines achieve small errors for temporal and in-plane spatial metrics but larger errors for out-of-plane joint angles~\cite{ref6}, so the 2D--3D gap is task- and joint-dependent rather than global, and 3D systems remain more resource-intensive and less accessible for routine clinical video~\cite{ref27}. Our design is therefore two-dimensional by intent: the dense silhouette captures the sustained postural and whole-body signals that dominate dystonia and chorea in our data, and the single-image 3D mesh is reserved for the setting where depth is mechanistically informative, out-of-plane posturing and axial rotation, as a hypothesis for future quantitative work rather than an assumed gain.

\subsection{Interpretability accounts for both successes and the failure mode}

A feature-level analysis showed phenomenology-specific, clinically coherent signals: chorea localized to the cranial segment and proximal upper limbs with whole body representation, athetosis to interior-point dispersion in the lower limbs and shoulder asymmetry, tremor to axial oscillation, and dystonia to sustained postural and height change. The same analysis explains the consistent failure of myoclonus, whose signature was weak and diffuse, with no region reaching a strong effect size, consistent with brief, jerk-like events being poorly summarized by 10-second window statistics and under-represented in training. This is a representational limitation, not a decision-layer one: a threshold can be recalibrated only where the backbone already encodes discriminative signal~\cite{ref33}. Because the failure lies in the temporal window rather than in a missing depth dimension, 3D features are unlikely to rescue it; short-window or event-based representations are the more promising route. This failure deserves to be stated against clinical experience and against the literature, because myoclonus is not a difficult sign at the bedside: an experienced clinician recognizes a brief, shock-like jerk immediately. The discrepancy is therefore informative about our representation rather than about the phenomenology. Computer-vision studies that have reported success on brief events did so with acquisition and analysis designed for them, for example myoclonus severity scored from recordings standardized to the Unified Myoclonus Rating Scale and analysed with pose and body-movement features on short segments~\cite{ref27}, or eye tics quantified frame by frame with head-pose normalization~\cite{ref30}. Both settings preserve the temporal resolution at which a jerk is defined, whereas our 10-second summary statistics average it away, and the examinations, especially in pediatric dataset\_2, were neither standardized nor designed to elicit or isolate myoclonus. We therefore read our result not as evidence that myoclonus is invisible to video analysis, but as evidence that detecting it requires event-level representations and dedicated tasks, which we did not have here.

\subsection{Calibration transfers reliably but is not reducible to a simple rule}

Transfer required adapting only the decision step, never the backbone which is a low-burden recipe needing a handful of annotated patients rather than full retraining~\cite{ref41,ref42}. We characterized this choice exhaustively: across 792 enumerated and 2,000 sampled calibration subsets, held-out performance was stable, and the \emph{a priori} clinician-selected subsets sat in the upper part of every distribution. Against a label-permutation null, that stability is informative in the pediatric dataset, whose distribution sits above the null, but not in the adult dataset, whose distribution coincides with it. Yet simple structural properties of a subset such as phenotypic coverage and mean inter-rater agreement predicted held-out performance only weakly and inconsistently, so calibration-set selection cannot be reduced to a coverage- or agreement-based rule; the practical recommendation is to follow the same \emph{a priori} clinical criteria used here.

\subsection{Clinical implications}

The decision rule was deliberately conservative, producing no false positives under the consensus-based definitions as the safer error profile for a screening or decision-support tool, which should avoid over-calling a phenomenology and triggering unnecessary work-up~\cite{ref18}. A conservative, reproducible, low-cost readout from routine video could support the assessment of complex, co-occurring movement disorders, and would be particularly consequential around DBS, where the observable phenotype is itself modulated by stimulation state and objective longitudinal phenotyping could inform programming and outcome assessment~\cite{ref49,ref50}. In rare and combined movement disorders, recognition of co-occurring phenomenologies can orient the syndromic hypothesis and, consequently, the genetic work-up; a reproducible and deliberately conservative multi-label readout derived from routine video could support this process within specialist care and facilitate remote case review through referral networks, in a field where diagnostic delay is often measured in years~\cite{ref5}.

\subsection{Limitations}

Several limitations bound these conclusions. The datasets are small (25 training; 12 pediatric, 20 adult, 2 standardized external) and the held-out sets modest (7 and 15 patients), so the findings are a proof of concept requiring prospective, multi-centre validation. Headline metrics rest on these small held-out sets and on tightened, agreement-based labels; we therefore report the full label-definition ladder and the complete calibration-subset distribution rather than a single figure, and per-phenomenology metrics for the rarest signs (three to seven positives) are unstable and reported as non-evaluable, with apparently high (ballismus, 0.80) or low (tics, 0.15) values being small-sample artefacts rather than interpretable estimates. The adult dataset was scored by only two raters, limiting the strength of its consensus labels relative to the five-rater pediatric dataset. Brief, jerk-like phenomenologies (myoclonus, tics) remain a failure mode, and low-prevalence phenomenologies could not be evaluated. Finally, inter-rater disagreement in this domain is itself part of the clinical signal rather than mere noise~\cite{ref13,ref27,ref29}, which both motivates the agreement-based analysis and bounds the achievable reference standard. The observed cross-dataset transfer should therefore be interpreted as evidence of robustness across the populations tested here rather than as proof of general transportability. Broad pretraining may facilitate transfer, but does not guarantee robustness under distribution shift; foundation-model-based systems still require external validation, local calibration, and continued performance monitoring~\cite{ref51}. Deployment also has a governance dimension specific to in-context classification: the model carries its context set with it, so a deployed instance ships 5,618 rows of per-window numerical descriptors and their labels, de-identified by construction but still patient-derived data.

\subsection{Future directions}

Natural extensions include quantitative 3D and multi-view kinematics for out-of-plane signs, dedicated short-window for tasks with optimal visibility or event-based representations for myoclonus and tics, and prospective multi-centre evaluation ensuring increase in the number of subjects and phenomenologies, diversity of data, with adjudicated labels. Coupling objective phenotype outputs to DBS workflows is the most immediate translational direction~\cite{ref49}. More speculatively, coupling phenotype outputs to a mechanistic, biologically structured circuitry model could allow movement-disorder phenotypes and related kinematic features to serve as a validation criterion for circuit-level hypotheses, linking observable phenomenology to underlying network perturbation. We regard this as a longer-term research direction rather than a near-term claim.

\section{Conclusion}

For the clinician, three points follow from this work. First, a phenotyping model built entirely from foundation models can be moved to a new centre without retraining: annotating a handful of patients to recalibrate the decision step was sufficient here, which is a realistic requirement for a clinical service. Second, the readout is most trustworthy exactly where it matters, on the signs that raters agree upon, and it is deliberately conservative, so a reported phenomenology is unlikely to be a false alarm; dystonia was detected near-perfectly and chorea and tremor were recovered through calibration alone. Third, the method is not yet complete: brief myoclonic jerks escape a representation built on 10-second windows, and this is a limitation of the representation rather than of the decision rule. These findings are a proof of concept on small datasets, not evidence of clinical readiness. With prospective multi-centre validation and event-level representations, a reusable backbone with light local calibration could provide reproducible, low-cost support for the assessment of complex, co-occurring movement disorders across the lifespan.

\section*{CRediT authorship contribution statement}
\begin{itemize}\setlength\itemsep{0pt}
  \item \emph{Conceptualization:} LC, DD, EM, G-MH, JB, GH, XV
  \item \emph{Investigation:} LC, ZS, DD, MC-J, JDO-E, MD, SH, CH, ND, GH
  \item \emph{Methodology:} LC, ZS, DD, MMUR, MC-J, JDO-E, G-MH, EMM, OO, XV
  \item \emph{Data curation:} LC, DD, ZS, MC-J, JDO-E, MD, SH, CH, ND, GH
  \item \emph{Visualization:} LC, ZS, MMUR, EMM, OO, XV
  \item \emph{Writing - original draft:} LC, ZS, DD, MMUR, XV
  \item \emph{Writing - review \& editing:} LC, ZS, DD, MC-J, MMUR, JDO-E, MD, SH, G-MH, CH, ND, EMM, JB, GH, OO, XV
  \item \emph{Supervision:} LC, XV
\end{itemize}

\section*{Data availability}
The data supporting the findings of this study are publicly available in the CODY record on Zenodo at \url{https://doi.org/10.5281/zenodo.22232609}.

The analysis code is publicly available on GitHub at \url{https://github.com/xaviervasques/cody-sam3}.

\section*{Declaration of competing interest}
The authors declare no conflicts of interest.

\section*{Acknowledgements}
This work was supported by Dystonia Medical Research Foundation, Canada (Grant 126616598 RR0001).

\section*{Conflicts of interest}
D.D. received honoraria for expert opinion from TEVA. J.B. is a share-holder of ONWARD Medical B.V., a company developing products for stimulation of the spinal cord, not related to this~research. All other authors declare no competing interests.

\clearpage
\appendix
\section*{Supplementary material}
\addcontentsline{toc}{section}{Supplementary material}
\setcounter{table}{0}
\renewcommand{\thetable}{S\arabic{table}}
\begin{table}[htbp]
\centering
\caption{\textbf{Feature-tier comparison: patient-level ROC-AUC under leave-one-patient-out cross-validation on the training dataset.} n+ = number of positive patients (supervised-signal layer); the retained Tier~2 column is shown in bold. Macro averages are reported over all eight phenomenologies.}
\label{tab:S1}
\small
\begin{tabular}{lcccc}
\toprule
\textbf{Phenomenology} & \textbf{n+} & \textbf{Tier 1 (494)} & \textbf{Tier 2 (1,539)} & \textbf{Tier 3 (6,346)} \\
\midrule
Dystonia & 21 & 0.86 & \textbf{0.88} & 0.91 \\
Tremor & 18 & 0.78 & \textbf{0.82} & 0.69 \\
Myoclonus & 17 & 0.32 & \textbf{0.42} & 0.39 \\
Athetosis & 12 & 0.90 & \textbf{0.80} & 0.70 \\
Chorea & 8 & 0.74 & \textbf{0.77} & 0.76 \\
Stereotypies & 7 & 0.44 & \textbf{0.50} & 0.48 \\
Ballismus & 3 & 0.91 & \textbf{0.80} & 0.47 \\
Tics & 3 & 0.46 & \textbf{0.15} & 0.00 \\
\midrule
\textbf{Macro (all 8)} & & \textbf{0.68} & \textbf{0.64} & \textbf{0.55} \\
\textbf{Macro (excluding stereotypies, ballismus, tics)} & & \textbf{0.72} & \textbf{0.74} & \textbf{0.69} \\
\bottomrule
\end{tabular}
\end{table}


\begin{thebibliography}{99}
\bibitem{ref1} Sanger TD, Chen D, Fehlings DL, Hallett M, Lang AE, Mink JW, et al. Definition and classification of hyperkinetic movements in childhood. Mov Disord 2010;25:1538--49. \url{https://doi.org/10.1002/mds.23088}.

\bibitem{ref2} Williams S, Relton SD, Fang H, Alty J, Qahwaji R, Graham CD, et al. Supervised classification of bradykinesia in Parkinson's disease from smartphone videos. Artif Intell Med 2020;110:101966. \url{https://doi.org/10.1016/j.artmed.2020.101966}.

\bibitem{ref3} Brooks C, Eden G, Chang A, Demanuele C, Kelley Erb M, Shaafi Kabiri N, et al. Quantification of discrete behavioral components of the MDS-UPDRS. J Clin Neurosci 2019;61:174--9. \url{https://doi.org/10.1016/j.jocn.2018.10.043}.

\bibitem{ref4} Postuma RB, Berg D, Stern M, Poewe W, Olanow CW, Oertel W, et al. MDS clinical diagnostic criteria for Parkinson's disease. Mov Disord 2015;30:1591--601. \url{https://doi.org/10.1002/mds.26424}.

\bibitem{ref5} Bertram KL, Williams DR. Delays to the diagnosis of cervical dystonia. J Clin Neurosci 2016;25:62--4. \url{https://doi.org/10.1016/j.jocn.2015.05.054}.

\bibitem{ref6} Tang W, Van Ooijen PMA, Sival DA, Maurits NM. Automatic two-dimensional \& three-dimensional video analysis with deep learning for movement disorders: A systematic review. Artif Intell Med 2024;156:102952. \url{https://doi.org/10.1016/j.artmed.2024.102952}.

\bibitem{ref7} Albanese A, Bhatia K, Bressman SB, DeLong MR, Fahn S, Fung VSC, et al. Phenomenology and classification of dystonia: A consensus update. Mov Disord 2013;28:863--73. \url{https://doi.org/10.1002/mds.25475}.

\bibitem{ref8} Mahlknecht P, Krismer F, Poewe W, Seppi K. Meta-analysis of dorsolateral nigral hyperintensity on magnetic resonance imaging as a marker for Parkinson's disease. Mov Disord 2017;32:619--23. \url{https://doi.org/10.1002/mds.26932}.

\bibitem{ref9} Comella CL, Leurgans S, Wuu J, Stebbins GT, Chmura T, and The Dystonia Study Group. Rating scales for dystonia: A multicenter assessment. Mov Disord 2003;18:303--12. \url{https://doi.org/10.1002/mds.10377}.

\bibitem{ref10} Burke RE, Fahn S, Marsden CD, Bressman SB, Moskowitz C, Friedman J. Validity and reliability of a rating scale for the primary torsion dystonias. Neurology 1985;35:73--73. \url{https://doi.org/10.1212/WNL.35.1.73}.

\bibitem{ref11} Sgandurra G, Olivieri I, Casarano M, Di Pietro R, Menici V, Velli C, et al. Inter and intra-rater reliability and minimal detectable difference of Movement Disorder-Childhood Rating Scale. Eur J Phys Rehabil Med 2018;54. \url{https://doi.org/10.23736/S1973-9087.17.04661-5}.

\bibitem{ref12} Sciacca G, Van Der Stouwe AMM, Centen LM, Tuitert I, Dalenberg JR, Svorenova T, et al. Next move in movement disorders (NEMO): the best clinical tasks for the visibility of essential tremor, dystonia, cortical myoclonus and myoclonus-dystonia. Parkinsonism Relat Disord 2025;138:107963. \url{https://doi.org/10.1016/j.parkreldis.2025.107963}.

\bibitem{ref13} Sadnicka A, Edwards MJ. Between Nothing and Everything: Phenomenology in Movement Disorders. Mov Disord 2023;38:1767--73. \url{https://doi.org/10.1002/mds.29584}.

\bibitem{ref14} Li MH, Mestre TA, Fox SH, Taati B. Vision-based assessment of parkinsonism and levodopa-induced dyskinesia with pose estimation. J NeuroEngineering Rehabil 2018;15:97. \url{https://doi.org/10.1186/s12984-018-0446-z}.

\bibitem{ref15} Zhang H, Ho ESL, Zhang FX, Del Din S, Shum HPH. Pose-based tremor type and level analysis for Parkinson's disease from video. Int J Comput Assist Radiol Surg 2024;19:831--40. \url{https://doi.org/10.1007/s11548-023-03052-4}.

\bibitem{ref16} Friedrich MU, Roenn A-J, Palmisano C, Alty J, Paschen S, Deuschl G, et al. Validation and application of computer vision algorithms for video-based tremor analysis. Npj Digit Med 2024;7:165. \url{https://doi.org/10.1038/s41746-024-01153-1}.

\bibitem{ref17} Di Biase L, Pecoraro PM, Bugamelli F. AI Video Analysis in Parkinson's Disease: A Systematic Review of the Most Accurate Computer Vision Tools for Diagnosis, Symptom Monitoring, and Therapy Management. Sensors 2025;25:6373. \url{https://doi.org/10.3390/s25206373}.

\bibitem{ref18} Friedrich MU, Relton S, Wong D, Alty J. Computer Vision in Clinical Neurology: A Review. JAMA Neurol 2025;82:407. \url{https://doi.org/10.1001/jamaneurol.2024.5326}.

\bibitem{ref19} Guo Z, Zeng W, Yu T, Xu Y, Xiao Y, Cao X, et al. Vision-Based Finger Tapping Test in Patients With Parkinson's Disease via Spatial-Temporal 3D Hand Pose Estimation. IEEE J Biomed Health Inform 2022;26:3848--59. \url{https://doi.org/10.1109/JBHI.2022.3162386}.

\bibitem{ref20} Morinan G, Dushin Y, Sarapata G, Rupprechter S, Peng Y, Girges C, et al. Computer vision quantification of whole-body Parkinsonian bradykinesia using a large multi-site population. Npj Park Dis 2023;9:10. \url{https://doi.org/10.1038/s41531-023-00454-8}.

\bibitem{ref21} Amprimo G, Masi G, Olmo G, Ferraris C. Deep Learning for hand tracking in Parkinson's Disease video-based assessment: Current and future perspectives. Artif Intell Med 2024;154:102914. \url{https://doi.org/10.1016/j.artmed.2024.102914}.

\bibitem{ref22} Cao Z, Hidalgo G, Simon T, Wei S-E, Sheikh Y. OpenPose: Realtime Multi-Person 2D Pose Estimation Using Part Affinity Fields. IEEE Trans Pattern Anal Mach Intell 2021;43:172--86. \url{https://doi.org/10.1109/TPAMI.2019.2929257}.

\bibitem{ref23} Mathis A, Mamidanna P, Cury KM, Abe T, Murthy VN, Mathis MW, et al. DeepLabCut: markerless pose estimation of user-defined body parts with deep learning. Nat Neurosci 2018;21:1281--9. \url{https://doi.org/10.1038/s41593-018-0209-y}.

\bibitem{ref24} Vakunov A, Chang C-L, Zhang F, Sung G, Grundmann M, Bazarevsky V. MediaPipe hands: On-device real-time hand tracking, 2020.

\bibitem{ref25} Brügge NS, Sallandt GM, Schappert R, Li F, Siekmann A, Grzegorzek M, et al. Automated Motor Tic Detection: A Machine Learning Approach. Mov Disord 2023;38:1327--35. \url{https://doi.org/10.1002/mds.29439}.

\bibitem{ref26} Tang Y, Béjar B, Essoe JK-Y, McGuire JF, Vidal R. Facial Tic Detection in Untrimmed Videos of Tourette Syndrome Patients 2022. \url{https://doi.org/10.48550/ARXIV.2211.03895}.

\bibitem{ref27} Pecoraro PM, Marsili L, Espay AJ, Bologna M, Di Biase L. Computer Vision Technologies in Movement Disorders: A Systematic Review. Mov Disord Clin Pract 2025;12:1229--43. \url{https://doi.org/10.1002/mdc3.70123}.

\bibitem{ref28} Vizcarra JA, Yarlagadda S, Xie K, Ellis CA, Spindler M, Hammer LH. Artificial Intelligence in the Diagnosis and Quantitative Phenotyping of Hyperkinetic Movement Disorders: A Systematic Review. J Clin Med 2024;13:7009. \url{https://doi.org/10.3390/jcm13237009}.

\bibitem{ref29} Martínez-García-Peña R, Koens LH, Azzopardi G, Tijssen MAJ. Video-Based Data-Driven Models for Diagnosing Movement Disorders: Review and Future Directions. Mov Disord 2025;40:2046--66. \url{https://doi.org/10.1002/mds.30327}.

\bibitem{ref30} Conelea C, Liang H, DuBois M, Raab B, Kellman M, Wellen B, et al. Automated Quantification of Eye Tics Using Computer Vision and Deep Learning Techniques. Mov Disord 2024;39:183--91. \url{https://doi.org/10.1002/mds.29593}.

\bibitem{ref31} Schappert R, Verrel J, Brügge NS, Li F, Paulus T, Becker L, et al. Automated Video-Based Approach for the Diagnosis of Tourette Syndrome. Mov Disord Clin Pract 2024;11:1136--40. \url{https://doi.org/10.1002/mdc3.14158}.

\bibitem{ref32} Peach R, Friedrich M, Fronemann L, Muthuraman M, Schreglmann SR, Zeller D, et al. Head movement dynamics in dystonia: a multi-centre retrospective study using visual perceptive deep learning. Npj Digit Med 2024;7:160. \url{https://doi.org/10.1038/s41746-024-01140-6}.

\bibitem{ref33} Hyppönen J, Hakala A, Annala K, Zhang H, Peltola J, Mervaala E, et al. Automatic assessment of the myoclonus severity from videos recorded according to standardized Unified Myoclonus Rating Scale protocol and using human pose and body movement analysis. Seizure 2020;76:72--8. \url{https://doi.org/10.1016/j.seizure.2020.01.014}.

\bibitem{ref34} Van Der Stouwe AMM, Tuitert I, Giotis I, Calon J, Gannamani R, Dalenberg JR, et al. Next move in movement disorders (NEMO): developing a computer-aided classification tool for hyperkinetic movement disorders. BMJ Open 2021;11:e055068. \url{https://doi.org/10.1136/bmjopen-2021-055068}.

\bibitem{ref35} Souei Z, Mushhood Ur Rehman M, Akram H, Bloch J, Chabardes S, Fasano A, et al. Artificial intelligence in deep brain stimulation for movement disorders: a systematic review and technology readiness assessment. Npj Digit Med 2026. \url{https://doi.org/10.1038/s41746-026-03015-4}.

\bibitem{ref36} Cif L, Demailly D, Horvàth GA, Escobar JDO, Dorison N, Jiménez MC, et al. Deep Learning Pose Estimation for Multi-Label Recognition of Combined Hyperkinetic Movement Disorders 2026. \url{https://doi.org/10.48550/ARXIV.2602.00163}.

\bibitem{ref37} Cif L, Demailly D, Souei Z, Rehman MMU, Escobar JDO, Jiménez MC, et al. Simultaneous hyperkinetic movement disorders phenotyping: a cross-cohort pediatric transfer study using routine videos, markerless pose estimation and a tabular foundation model 2026. \url{https://doi.org/10.48550/ARXIV.2606.07674}.

\bibitem{ref38} Carion N, Gustafson L, Hu Y-T, Debnath S, Hu R, Suris D, et al. SAM 3: Segment Anything with Concepts 2025. \url{https://doi.org/10.48550/ARXIV.2511.16719}.

\bibitem{ref39} Ma J, He Y, Li F, Han L, You C, Wang B. Segment anything in medical images. Nat Commun 2024;15:654. \url{https://doi.org/10.1038/s41467-024-44824-z}.

\bibitem{ref40} Hollmann N, Müller S, Purucker L, Krishnakumar A, Körfer M, Hoo SB, et al. Accurate predictions on small data with a tabular foundation model. Nature 2025;637:319--26. \url{https://doi.org/10.1038/s41586-024-08328-6}.

\bibitem{ref41} Qu J, Holzmüller D, Varoquaux G, Morvan ML. TabICL: A Tabular Foundation Model for In-Context Learning on Large Data 2025. \url{https://doi.org/10.48550/ARXIV.2502.05564}.

\bibitem{ref42} Qu J, Holzmüller D, Varoquaux G, Morvan ML. TabICLv2: A better, faster, scalable, and open tabular foundation model 2026. \url{https://doi.org/10.48550/ARXIV.2602.11139}.

\bibitem{ref43} Cif L, Demailly D, Horvàth GA, Ortigoza Escobar JD, Dorison N, Castro Jiménez M, et al. Deep Learning Pose Estimation for Phenotyping of Co-Occurring Hyperkinetic Movement Disorders. Ann Clin Transl Neurol 2026:acn3.70474. \url{https://doi.org/10.1002/acn3.70474}.

\bibitem{ref44} Yang X, Kukreja D, Pinkus D, Sagar A, Fan T, Park J, et al. SAM 3D Body: Robust Full-Body Human Mesh Recovery 2026. \url{https://doi.org/10.48550/ARXIV.2602.15989}.

\bibitem{ref45} Jocher G, Chaurasia A, Qiu J. Ultralytics yolov8 2023.

\bibitem{ref46} Tang W, Van Ooijen PMA, Sival DA, Maurits NM. 2D Gait Skeleton Data Normalization for Quantitative Assessment of Movement Disorders from Freehand Single Camera Video Recordings. Sensors 2022;22:4245. \url{https://doi.org/10.3390/s22114245}.

\bibitem{ref47} Mifsud J, Embry KR, Macaluso R, Lonini L, Cotton RJ, Simuni T, et al. Detecting the symptoms of Parkinson's disease with non-standard video. J NeuroEngineering Rehabil 2024;21:72. \url{https://doi.org/10.1186/s12984-024-01362-5}.

\bibitem{ref48} Hoang T-H, Zehni M, Xu H, Heintz G, Zallek C, Do MN. Towards a Comprehensive Solution for a Vision-Based Digitized Neurological Examination. IEEE J Biomed Health Inform 2022;26:4020--31. \url{https://doi.org/10.1109/JBHI.2022.3167927}.

\bibitem{ref49} Dixon TC, Strandquist G, Zeng A, Frączek T, Bechtold R, Lawrence D, et al. Movement-responsive deep brain stimulation for Parkinson's disease using a remotely optimized neural decoder. Nat Biomed Eng 2025;10:110--24. \url{https://doi.org/10.1038/s41551-025-01438-0}.

\bibitem{ref50} Tisch S, Kumar KR. Pallidal Deep Brain Stimulation for Monogenic Dystonia: The Effect of Gene on Outcome. Front Neurol 2021;11:630391. \url{https://doi.org/10.3389/fneur.2020.630391}.

\bibitem{ref51} Harris C, Schmidgall S, Rapuri S, Hwang K, Moor M, Stevens RD. A perspective on foundation models in intensive care medicine. Artif Intell Med 2026;181:103515. \url{https://doi.org/10.1016/j.artmed.2026.103515}.
\end{thebibliography}
\end{document}